\documentstyle[psfig]{l-aa-dipastro}
	\def\msol{M$_\odot$}

	\def\p1{\partial}

\def\smallskip{\vskip 8pt}

\begin{document}

\thesaurus{ }

\title{Evolution of Massive Stars under  New Mass-Loss Rates for 
RSG: Is the mystery of the missing blue gap solved ?}

\author{Bernardo Salasnich$^1$, Alessandro Bressan$^2$, and 
           Cesare Chiosi$^{3,1}$ }
\institute{
    $^1$ Department of Astronomy, University of Padova,
      Vicolo dell'Osservatorio 5, 35122 Padova, Italy \\         
    $^2$ Astronomical Observatory, 
      Vicolo dell'Osservatorio 5, 35122 Padova, Italy\\ 
    $^3$ European Southern Observatory, K-Schwarzschild-strasse 2,
      D-85748, Garching bei Muenchen, Germany }

\offprints{A. Bressan  }

\date{Received: November 1997;  Accepted: }

\maketitle
\markboth{B. Salasnich, A. Bressan, \& C. Chiosi: Diffusion and 
                new mass-loss rates of  RSG stars }{}

\begin{abstract} 
In this paper we present new models of massive stars
based on recent advancements in the theory of  diffusive mixing and a
new empirical formulation of the mass-loss rates of red supergiant
stars. We synthesize, by means of a simple diffusive algorithm, the
results of  complex studies  on non local convection
(overshooting region) by Xiong (1989) and Grossman
(1996) and compute two sets of stellar models of
massive stars  with initial chemical composition [Z=0.008,  Y=0.25]
and [Z=0.020, Y=0.28]. Mass loss by stellar wind is also taken into
account according to the empirical relationship by de Jager et al. (1988).
Stars with initial mass in the range 6 to 120 \msol\ are
followed from the zero age main sequence till core He-exhaustion. 
Particular attention is paid to the 20
\msol\ star as the prototype of the evolution of massive stars in 
the luminosity (mass) interval in which both blue and red supergiants
occur in the HR diagram (HRD).  The  models
confirm that, in the evolution of a  massive star with
mass loss, the dimension of the H-exhausted core and the
efficiency of intermediate mixing strongly affect the evolution during 
the subsequent core He-burning phase, the extension of the blue loops 
in particular. However, despite the new mixing prescription,
also these models share the same problems of older models in literature
as far as the interpretation of the observational distribution of stars
across the HRD is concerned. Examining possible causes of the failure,
we find that the rate of mass loss for the red supergiant stages implied 
by the de Jager et al. (1988) relationship under-estimates the 
observational values by a large factor. Revising the whole problem, first
we adopt the recent formulation by Feast (1991) based on infrared data,
and secondly we take also into account the possibility that the dust to gas 
ratio varies with the stellar luminosity (as suggested by the observations).
Stellar models are then calculated with the new prescription for the 
mass-loss rates during the red supergiant stages in addition to the new
diffusive algorithm. The  models now possess very extended loops in the HRD 
 and  are able to match the distribution of stars across the HRD from the 
earliest to the latest spectral types both in 
the Milky Way, LMC and SMC.  During the
loop phase the surface abundance of helium is about 0.5, in good agreement 
with the enhancement of this element observed in blue supergiant stars.
Finally, because the surface H-abundance is close to the
limit adopted to start the Wolf-Rayet phase (WNL type),  we suggest
that a new channel is possible for the formation of low
luminosity WNL stars, i.e. by progenitors whose mass can be as low as
20 \msol, that have evolved horizontally across the HRD following the
blue-red-blue scheme and suffering large mass loss during the red supergiant
stages.


\end{abstract}

\section {Introduction}

Among the physical phenomena  at the base of stellar
models,  convective mixing is perhaps the most difficult one to
handle properly, even if is likely to play the  dominant role.
 Suffice it to recall the long debate about the choice of the
stability criterion and the related   phenomena of
semiconvection and overshoot (Schwarzschild \& Harm 1965). Convection
theory grows more and more complicated, 
(Xiong 1979, 1984; 
Canuto \& Mazzitelli 1991; Canuto 1992; Grossmann \& Narayan 1993; 
Canuto 1996; Grossmann \& Taam 1996),
but its application to the stellar models  still requires a great
deal of over-simplification (Zahn 1991; Alongi et al. 1993; Deng 1993;
Deng et al. 1996a,b). 

Regarding to  the structure and evolution of massive stars, the 
impressive body of literature on
the mixing schemes  shows that 
  mixing  in stellar interiors is a complex phenomenon because different
physical instabilities and mixing  processes arise in
different regions of a star.  The reader is referred to  Chiosi et al. 
(1992)  for an exhaustive review
of the subject. It is worth recalling here that, following core
H-exhaustion, the possible occurrence of   
an intermediate convective layer   and
consequent modification of the chemical profile over there 
 are strictly
related to the adopted stability criterion, either Schwarzschild 
($\nabla_{r} < \nabla_{a}$) or Ledoux 
 ($\nabla_{r} < \nabla_{a}+ \nabla_{\mu}$), with profound
consequences for the subsequent core  He-burning phase.

Another phenomenon playing  a dominant role in the
evolution of  massive stars is mass-loss by stellar wind. In early type
stars, the radiation driven wind theory developed starting from the seminal 
studies of Lucy \& Solomon (1970) and Castor et al. (1975), followed by
 Abbott (1982),
Pauldrach et al. (1986), Owocki et al. (1988) and many others, 
until the magistral review 
article by Kudritzski (1997, and
references) provides quantitative predictions for the mass-loss
rates that find confirmation in the observational data. 
There are still  several aspects  to be clarified, for instance the 
very high ratio $v_{\infty} / Lc^{-1}$ observed in some Wolf-Rayet 
stars (cf. Hillier 1996 and references). The situation is much less
settled with the late type supergiant stars, cf. the  review by 
Lafon \& Berruyer (1991), and much  work  on this subject refers
to bright AGB stars. The most likely mechanism
is radiation pressure on dust grains (cf. Kwok 1975; Gail \& Sedlmayr
1987). The situation is further complicated by possible effects due
to pulsation (cf. Willson 1988), and perhaps sound waves generated either by
convection in the mantle of the stars or by pulsation at high eigenmodes
(cf. Pijpers \& Hearn 1989; Pijpers \& Habing 1989).

In star more massive than say  30 \msol, the
evolution is almost fully  determined  by  mass loss starting from  the
H-burning phase (see. Chiosi \& Maeder 1986).
In stars of about  20 \msol, both internal mixing and mass-loss 
affect the evolutionary path  in the HR diagram (HRD), 
so that stars in this mass range
are a workbench for testing theories of massive star structure and evolution,
thanks also to the lucky circumstance that in the luminosity interval 
pertinent to a typical 20 \msol\  star counts in the Solar Vicinity 
and in the Magellanic Clouds are fairly complete (cf. Massey 1997). 

The comparison of extant theoretical models with observations
indicates,
however,  severe points of disagreement, probably due to our poor 
knowledge of the
above physical phenomena. The reader is referred to Chiosi (1997) for an
exhaustive discussion of these topics and  referencing. In brief, 
the distribution of stars
across the HRD in the luminosity range $\rm -7 \geq M_{bol} \geq -9$
and the surface chemical  abundances of these stars hint 
that after central H-exhaustion, a star should ignite core He-burning 
as a red supergiant (RSG), perform an extended blue loop up to the 
main sequence band, and eventually return to the RSG region, 
thus following the classical case A evolutionary scheme of Chiosi \&
Summa (1970). If we
take the progenitor of SN1987a as an indicator of the final fate of
the evolution of massive single stars, then the model should perform
a final loop toward hotter temperatures during the central C-burning 
phase. No stellar model is found in literature able to fit
this simple evolutionary scheme. In particular,  the extension of the
loop into the blue region is far too short with respect to what
required by observations, and gives rise to the so called ``Blue
Hertzsprung Gap (BHG) problem''. In fact, the theory predicts the existence 
of a gap between the core H- and He-burning regions caused by the 
long lifetime
of the major nuclear phases and the short timescale  passing from 
the first  to the second one, so that very few stars
 should fall in the gap. In  
reality the opposite occurs and the vast majority of blue 
supergiants crowd the {\it forbidden} area.  
 Several ways out have been proposed in the past none of
which is really able to unravel the  mystery (cf. Chiosi 1997).

Furthermore, the predicted surface abundance of helium (and other elements as
well) are too low as compared to the observational data (cf. Herrero et al.
1992, Venn 1995, Lennon et al. 1996).

Finally, there is the long debated problem of the origin
of low luminosity WR stars. They are found in a region 
near the main sequence down to the luminosity  of a
 typical  15 \msol\  star. Low luminosity WR stars are
currently understood as the
descendants of the most massive stars  that under the
effects of a vigorous stellar wind lose their entire H-envelope and
reach such  low
luminosities  following the mass-luminosity relation 
(cf.  Maeder \& Conti  1994).
However, first of all the predicted effective temperature of WR stars is
by far  higher than observed, and secondly the lifetime during
which the evolutionary tracks populate the ``observed'' region is by
far too short. Attempts to cure the disagreement were not particularly
satisfactory.

In this paper we thoroughly investigate the effects of mixing and
mass-loss on the structure and evolution of a typical 20 \msol\
star in order to cast light on the coupling of the two phenomena in
this mass range. 

Mixing is described by means of a diffusive algorithm easily
applicable to stellar models. In Section 2 we present our 
formulation of diffusion, and in Section 3 we discuss our derivation of
the diffusion coefficient, which takes different values throughout
 a star.  We
describe the results obtained for our prototype  model of 20
\msol\ with particular emphasis on the value of the adopted
parameters. Two complete sets of stellar models with chemical
composition [Z=0.020, Y=0.280] and [Z=0.008, Y=0.250] are
computed and compared with observations,  highlighting the points
of uncertainty and disagreement that strongly hint for a new formulation of
the mass-loss rate during the RSG phase. The new rates of mass loss
are presented in Section 4 and are compared to the standard
ones.  In Section 5 we describe the results obtained with
the revised mass-loss 
rates, and show how the new stellar models are able to improve sensitively 
upon the solution  of the 
BHG problem. In Section 6 we thoroughly examine to which extent internal 
mixing and mass loss from RSG stage concur to determine the final result.
It emerges that mass loss is the dominant factor.
 Finally, some concluding remarks are drawn in Section 7.

\section {Mixing as a diffusive process } 

Several authors
have already made use of a diffusion approach to deal with
 mixing in stellar interiors
(Weaver \& Woosley 1978; Langer 1983; Langer et al. 1985; Deng 1993; 
Deng et al. 1996a; Gabriel 1995; Grossmann \& Taam 1996; 
Herwig et al. 1997
and references therein). The advantage with the diffusive approach
is that  the finite time-scale of the 
mixing process is naturally taken into account. In fact,
laboratory fluid-dynamics experiments and numerical simulations 
(cf. Canuto 1994 and references therein) 
show
that mixing is a rather slow process that becomes complete only at
asymptotic regimes. In a turbulent field the speed of mixing is not
high enough to wipe out inhomogeneities over the time scale of 
large scale motions. Recent  simulations (Freytag et al. 1996) show that
the time needed for complete mixing of a certain  volume is
already much longer than the simulation time.

However, as 
the nuclear time-scale is much longer than the characteristic turnover
time predicted by  the mixing length theory (MLT),   in principle the
unstable regions in the deep interiors of a star should be completely
homogenized over a nuclear time-scale. For this reason an unstable
region is usually homogenized without any particular  algorithm. 

Nevertheless, situations are met  in which this may lead to ambiguous
results, regarding to   massive stars in particular. In fact, during 
the core
H-burning phase, at the boundary of the convective
core a region develops  where matter tends to acquire a neutral condition with
respect to the instability criterion. Irrespectively of the 
stability criterion adopted,  a region with a gradient in chemical
composition can set in over the nuclear
time-scale, indicating that mixing cannot  be fully efficient over there.
After central H-burning, a convective zone develops inside the
region with a gradient in  chemical abundances.

 Observations seem to hint  that mixing should not occur in this region, 
at least while the star is crossing the HRD to enter the red supergiant 
phase. In fact the crossing time is much shorter than 
the nuclear lifetime, and comparable to the growing
time of a perturbation inside the region with a gradient in the
chemical abundance.
Once more, mixing must operate with a much smaller efficiency than that
derived on the basis of the turnover lifetime of  convective
eddies. 

Finally, there are numerical simulations of mixing inside the
overshoot region (Xiong 1989, Grossman $\&$ Narayan 1993, Freytag et al. 1996) 
suggesting that characteristic turbulent
velocities decay almost exponentially with the distance from the
unstable region, leading to an incomplete mixing there.

The diffusive description of the mixing process naturally accounts for the
different time-scales encountered in different circumstances,
provided that  suitable laws for the diffusion coefficient $D$ are specified.
The diffusion equation is

\begin{equation}
{{dX}\over {dt}}=\left( {{\p1 X}\over{\p1 t}}\right)_{nucl} +
{{\p1}\over{\p1 m_r}}
\left[ {(4\pi r^2\rho )^2D}{{\p1 X}\over{\p1 m_r}}\right]
\label{equazionediffusione}
\end{equation}

\noindent
where the term describing  the variation of the chemical abundance in
a mesh point $j$ at time $t$ due to nuclear burning has been
written explicitly.
The corresponding time-scale of the mixing process over  a distance $L$ is 

\begin{equation}
t_{diff}=L^{2}/D
\label{tdiff}
\end{equation}

\noindent
The use  of the diffusive algorithm in stellar model calculations 
offers two advantages: (i) Since mixing is a time-dependent process, 
the diffusive scheme takes into account the possibility of 
incomplete  mixing  over the
time-step $t_{evol}$ between two successive models. 
The region will be homogenized over  a time interval much longer than  the
diffusive time-scale. (ii) Adopting a unique
prescription, we can deal with different physical processes of
mixing such as full convection, semiconvection, rotationally induced
mixing, convective overshoot, that can occur in different regions of
a star.

\begin{figure}
\psfig{file=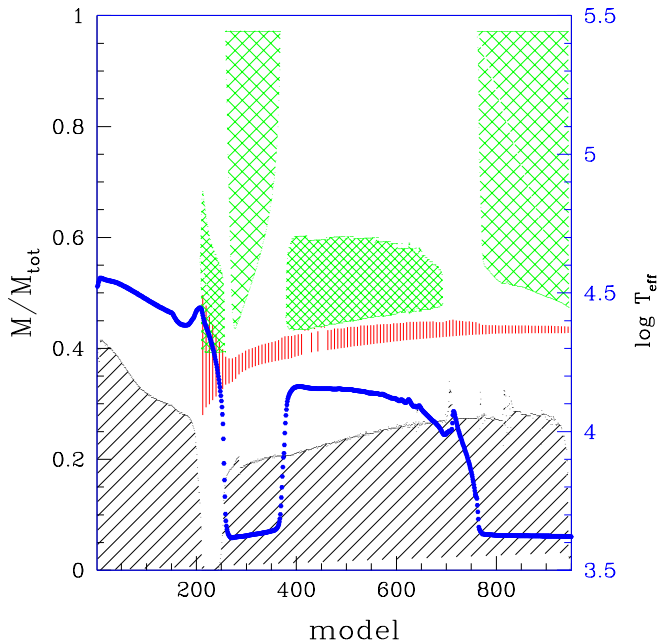,width=8.8cm}
\caption{Internal structure of the 20 \msol\ star with
chemical composition [Z=0.008, Y=0.250] as a function of the model. 
The single hatched
regions correspond to the convective core defined by the
Schwarzschild  criterion, whereas the double hatched zones show the
intermediate fully convective layers  and  the external convective 
envelope. The
vertical lines display the extension of the H-burning shell. The
boundaries are taken where the rate of nuclear energy release drops
below $10^{-3}$ of the   peak value. For the sake of clarity the
overshoot regions around the core are not drawn, likewise for  the
overshoot region at the base of the external envelope and at the
edges of the intermediate convective shell. Finally, the heavy dots show
the run of the effective temperature in the course of evolution (right
vertical axis).}
\label{ombra}
\end{figure}

\subsection{The diffusion coefficient $D$}

As already mentioned, different kinds of instability may arise in
different regions of a star in the course of its evolution, each of 
which requires a different prescription for $D$.

 For the sake of illustration, 
we shortly describe here the typical structure of a massive star
from the main sequence to late core He-burning, see Fig.~1,  highlighting the 
possible sources of instability and mixing that can occur.

During the H-burning phase   the star is composed by  a 
convectively unstable  core (the single hatched region)
surrounded by  formally stable radiative layers. Owing to the inertia
of the convective elements, a significant fraction of the
radiative region can be partially or totally mixed with the core,
the so-called {\it overshooting}.  As the  evolution
proceeds, above the H-burning core
 potentially unstable  oscillatory convection may develop 
(Merryfield 1995),  and eventually turn into
{\it semiconvection} over extended regions, across  which a suitable chemical 
profile is built up.
In the same layers, at the beginning of shell H-burning, one or more
fully convective zones may arise, depending on  the adopted stability
criterion (dense double-hatched areas in Fig.~1). These intermediate
convective regions can even 
penetrate into the H-burning shell (vertical lines in Fig.~1). The whole
structure is finally surrounded by an outer radiative (during the BSG
stages) or convective (during the RSG stages) envelope 
(double hatched zones in Fig.~1). 

Following Deng et al. (1996a), we distinguish three main regions of a star,
in which the treatment of  mixing requires different
prescriptions for the diffusion coefficient owing to the different
physical nature of the process inducing mixing.\\

\noindent
(a) {\it The homogeneous central region}

\noindent
With this  we mean the central core of the star which is 
unstable to convection according to the Schwarzschild criterion,
where the
convective elements have the same probability of crossing the nuclear
burning zone. So, no specific algorithm for
mixing is  required to make this region homogeneous (cf. Deng 1993).
In any case, the characteristic time of the homogenization process is
of the order of the convective turnover time. In principle, this
region should correspond to the very central sphere with radius equal
to the mean free path of the elements. However, we  will consider the
 whole central region unstable to convection as being 
homogenized over the nuclear time-scale, and
hence  adopt a diffusion coefficient able to secure  complete
mixing, i.e.

\begin{equation}
 D={1 \over 3} (v\times L_{c})
\label{d_cent}
\end{equation} 

\noindent
where $v$ is the turbulent velocity calculated with the MLT and
$L_{c}$ is the radius of the central unstable region. This is the
expression adopted by Weaver et al. (1978), Langer et al.
(1983, 1985), Deng (1993) and 
Deng et al. (1996a).\\

\noindent
(b)  {\it The overshoot region}

\noindent
Hydrodynamical modelling of stellar convection by Grossmann
(1996) shows
that turbulent velocities may penetrate into the surrounding stable
region with a decay scale which is of the order of one pressure scale
height $H_P$ [see also Grossman et al. (1993) for details].

\begin{figure}
\psfig{file=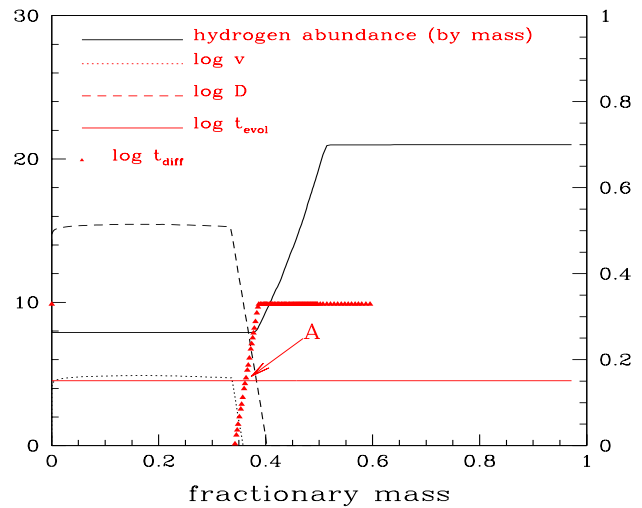,width=8.8cm,height=7.cm}
\caption {Time-scales involved during central H-burning as a function
of the fractional mass of the star. For a given model, diffusion is
efficient up to point A, where the diffusive time-scale (triangles) is
smaller than the evolutionary time-scale (thin solid line) between two
successive models (left scale in units of $log$ years). In this
example, the large value of $D$ ($\rm \approx 10^{15} cm^{2}/sec$, left 
vertical axis) ensures the complete homogenization of the core. The right
axis refers to the profile of hydrogen abundance (thick solid 
line).}
\label{strategy}
\end{figure}

Such a large extension of the overshoot region is consistent with the
results by Xiong (1985), who  showed that 
different physical quantities have different distance of
penetration into the radiative region, and that their fluctuations
decrease exponentially from the  Schwarzschild border. In this regard,
 see  Figs.~5a,b in Xiong (1985), Fig.~3 in Xiong (1989),
and the 60 \msol\ star in Xiong (1985), in which an
e-folding distance of $1.4 c_1  H_{P}$ is found, where $c_1$ is the
 efficiency
parameter of the  energy transport.
Assuming $c_1 =0.5$ like in Xiong (1985), we obtain an e-folding
distance of $0.7 H_{P}$, which is comparable with the value suggested by
Grossmann (1996). Recent hydrodynamical simulations
 by Freytag et al. (1996) reveal that the
transition from convective and hence mixed regions  to those not
affected by mixing  is
rather sharp due to the fast decline of the convective velocities, and
confirm moreover that  the velocity field continues beyond the region with
significant convective flux, declining exponentially with the depth.
Simulations of convective velocity field at the surface  of a
star (Freytag et al. 1996) suggest that the diffusion coefficient varies
with the distance $r$ from the border of the convective region according to

\begin{equation} 
 D= H_{P} v_0 \times exp[-2r/(H_v)]  
\end{equation}

\noindent
where $H_{v}$ and $v_{0}$ are the velocity scale height and velocity, 
respectively, at border of the convective zone. Furthermore, 
 $H_{v}$ is found to vary from $0.25 \pm 0.05 H_{P}$ in models of
A-type stars to
$1.0 \pm 0.1 H_{P}$ in models of white dwarfs, 
 thus indicating a significant dependence
of the diffusion coefficient upon the type of star.

The above mentioned studies suggest that 
beyond the convective region
turbulent velocities 
neither  vanish abruptly (no overshoot)
nor generate a fully mixed region (the standard picture of 
convective overshoot), but decay
exponentially into the stable layers.

On the basis of these considerations, we assume 
that in the overshoot regions  the diffusion coefficient 
declines exponentially with the
distance from the  border of the core, and adopt the pressure scale
height as the critical distance over which mixing is effective:

\begin{equation}
D=  {1 \over 3} H_{P} v_0 \times exp[-r/(\alpha_1 H_P)]
\label{d_ov}
\end{equation}

\noindent
with obvious meaning of the symbols. We
introduce the parameter $\alpha_{1}$ because the original formulation
by Xiong (1989) and Grossman (1996) (corresponding to $\alpha_1$ in
the range 0.5 to 2) yields complete homogenization over  too wide a
region around the unstable core, and the resulting stellar models are
unable to match most of the observational data (see the entries of
 Table~1 below).

This can be easily checked without calculating complete stellar models. 
Adopting the e-folding scale
deduced by Grossmann (1996)  and Xiong (1989),
and starting from  $D \sim 10^{15}$ $\rm cm^{2}~sec^{-1}$ at the
border of the unstable region, we obtain   $D \sim 3-5~ 10^{14}$ 
$\rm cm^{2}~sec^{-1}$ at one pressure scale height above the Schwarzschild
border. The corresponding diffusive time-scale within a region of thickness
 $H_{P}$ is
about 3 days. When compared to the nuclear time-scale this implies
complete homogenization of the matter up to the distance  $H_{p}$ above the
Schwarzschild core. In other words, the diffusion time-scale parallels
the nuclear lifetime at about  $4\times H_{P}$ above the unstable region. In
Fig.~2  we show the time-scales in question  during the core H-burning as
a function of the fractionary mass of the star. For a given model,
diffusion is efficient up to point A, where the diffusive time-scale
(triangles) is smaller than the evolutionary time-scale (solid 
line) between two successive models (all time-scales are expressed
in years on a logarithmic scale).
In this example, with $\alpha_{1} = 0.009$, the large value
of $D$ ($\rm \approx 10^{15} cm^{2}~sec^{-1}$, left vertical axis in Fig.~2)
ensures the complete homogenization of the core. The right vertical
axis  shows the profile of  hydrogen abundance (thick solid line). \\

\noindent
(c) {\it The intermediate convective region}

Following central H-exhaustion an extended region with a gradient in
molecular weight develops in which a convective zone may arise owing
to the so-called {\it oscillatory convection or overstability}. The
physics of this phenomenon has been studied by many authors by means of 
linear-stability analyses. 
In particular, Kato (1966) showed 
that, due to heat dissipation processes, in a region which is stable
against convection according to the Ledoux criterion but not to the
Schwarzschild criterion, infinitesimal perturbations grow on a
time-scale of the order of the thermal diffusion time-scale $t_{heat}$,
giving rise to oscillatory convection which  eventually mixes the
whole region.

It is thus conceivable that the diffusion coefficient in this region
is governed by  the thermal diffusion time-scale $t_{heat}$, which 
according to Langer et al. (1983, 1985) can be expressed as 

\begin{equation}
 t_{heat} = \frac{\varrho c_{p}}{K k^{2}} 
\end{equation}

\noindent 
where  

\begin{equation}
K =  \frac{4acT^{3}}{\chi \varrho}
\end{equation} 

\noindent
is the radiative-conductivity
coefficient, and $k^{-1}$ is the characteristic  scale-length of the
perturbations, approximated  here to $H_{P}$. 
All other symbols have their usual meaning.

Since  we do not know when  the perturbation  becomes finite and
what the real time-scale of the  mixing process is, we express
the diffusion coefficient as

\begin{equation}
D=L^{2}/t_{growth}
\label{d_int}
\label{dfsicz}
\end{equation}

\noindent
where $L$ is the size of the intermediate convective zones  as
defined by  the Schwarzschild criterion,  and $t_{growth} =
\alpha_{2} \times t_{heat}$, where $\alpha_{2}$ is a suitable constant
to be determined by comparing theoretical results with observations.

With these assumptions it turns out that
the diffusion coefficient $D$ is weakly dependent on the size $L$ of
the intermediate convective zone, contrary to Langer et al. (1983, 1985) 
and  Grossmann \& Taam (1996).

Complete mixing is ensured if $t_{growth} << t_{evol}$ (the
characteristic evolutionary time-scale).  In our modelling,  the whole
process is controlled by the parameter $\alpha_{2}$. Figs.~ 3 and 4 
show the
hydrogen content in  four successive models during the Kelvin-Helmoltz
phase, in which two extreme values for $\alpha_{2}$ have been adopted,
i.e. $\alpha_{2}=1000$ and $\alpha_{2}=0.5$. To better understand 
the effect of  diffusion  in this phase on the chemical profile
generated in previous  stages,  we introduced a
scalar quantity at the center of the unstable region and follow its
evolution in the four models. This is shown in the upper panels of
 Figs.~3 and  4.
These experiments are meant
to illustrate the profile generated by
diffusive mixing alone.

\begin{figure}[h]
\psfig{file=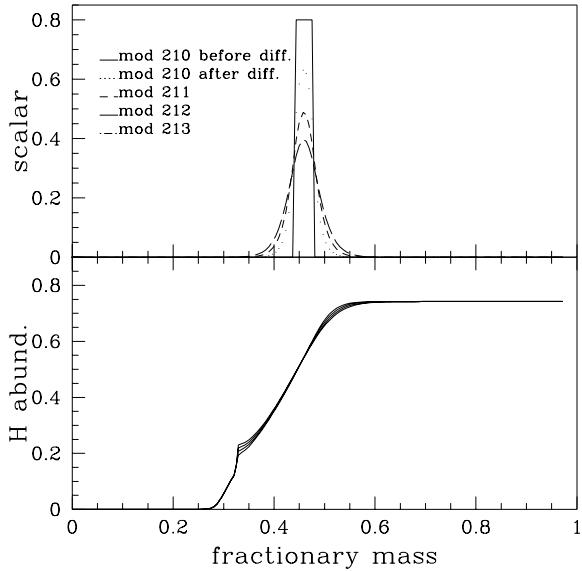,width=8.cm}
\caption { Hydrogen abundance by mass (bottom panel) as a function of
the fractionary mass of 4 successive models during the
Kelvin-Helmoltz phase. The upper panel shows the diffusion of a
scalar quantity arbitrarily introduced at the center of the
intermediate convective region of the first model. The
models are calculated with $\alpha_{1}=0.009$ and $\alpha_{2}=1000$
respectively.}
\label{scalare1000}
\end{figure}

\begin{figure}[h]
\psfig{file=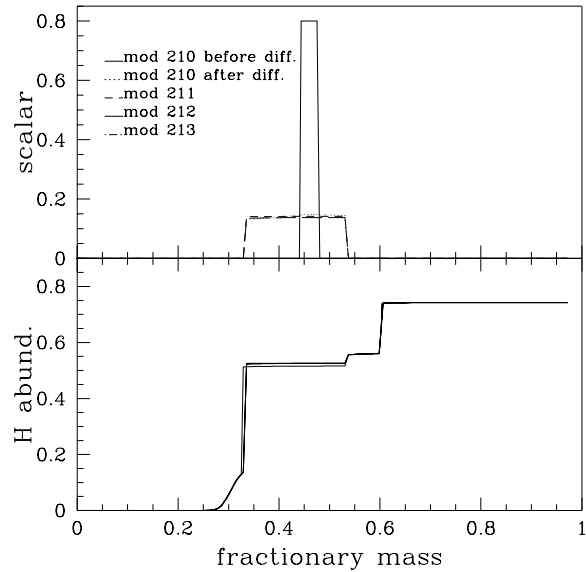,width=8.cm}
\caption {Same as Fig.~4,  but for  $\alpha_{1}=0.009$ and
$\alpha_{2}=0.5$}
\label{scalare05}
\end{figure}

It is worth noticing that by varying  $\alpha_{2}$, the chemical profiles
corresponding to the 
Schwarzschild (rectangular profile)  and the   Ledoux neutrality condition
(smoothed profile) are recovered.\\

\noindent
{\it Summary of the prescriptions for $D$}

\noindent
We have presented a simple diffusive scheme suited to dealing with 
 mixing  in different convective regimes.

In the central unstable region the adopted diffusion coefficient (eq.
\ref{d_cent}) secures complete homogenization on the nuclear time-scale.

In the overshoot region the diffusion coefficient decays
exponentially from the border of the Schwarzschild core, with a
typical scale-length equal to $\alpha_1 \times H_P$. This choice is
based on  the results of recent hydrodynamical studies showing that
convective turbulent velocities decay exponentially into the
surrounding stable radiative region. Notice that instead of
specifying a characteristic  length of the overshoot region we
specify the  scale-length of the decay of the overshoot process,
which sounds to be more  physically grounded. Since  current 
theoretical predictions  seem to over-estimate  the decay length,
we introduce a parameter
$\alpha_1$ to be  eventually  calibrated on the observations.

In the intermediate convective region which may develop 
in layers with a gradient in chemical abundances
(built up in previous stages according to the  Schwarzschild
criterion),  we adopt a diffusion
coefficient based on a fraction of the thermal dissipation time-scale
($t_{growth} = \alpha_{2} \times t_{heat}$). By changing the parameter
$\alpha_2$ we may recover the   Schwarzschild or the Ledoux criterion
for convective stability. This diffusive scheme 
secures  that   complete mixing occurs only when $t_{growth} <<
t_{evol}$.

\section{Stellar models with the new mixing scheme}

In this section we present stellar models calculated with the 
  above prescriptions for internal mixing. 
Prior to this,  we  perform a detailed study of  a typical  20 \msol\ 
star in order to check  the
response of the stellar structure  to  variations of the parameters
$\alpha_{1}$ and  $\alpha_{2}$. 
We adopt the chemical composition [Z=0.008,  Y=0.25], which is most
suited to stars in the LMC.
The evolution  is
followed from the main sequence to the core He- exhaustion stage.

In order to proceed further, we first need to specify the prescription
adopted for the mass-loss rate throughout the various evolutionary
phases. All other details concerning the input physics (opacity,
nuclear reaction rates, neutrino energy losses, etc. ) are as in
Fagotto (1994), to whom the reader is referred.

\begin{table*}
\caption{Stellar models with the new prescription for the diffusive 
mixing, standard mass-loss rates,  and  chemical composition 
[Z=0.008,  Y=0.25]. See the text for more details.}
\begin{scriptsize}
\begin{center}
\begin{tabular*}{180mm}{lc|cc|cc|cc|ccc|ccc|c}
\hline
\hline
\multicolumn{2}{c}{}  & \multicolumn {2}{c}{A} & \multicolumn
{2}{c}{B} &
\multicolumn{2}{c}{LOOP} & \multicolumn {6}{c}{} \\
$\alpha_{1}$ &$\alpha_{2}$&  $logL$ & $logT_{eff}$ & $logL$ &
$logT_{eff}$
& $logL$ & $logT_{eff}$ 
& $Q_{HE}$ & $Q_{XP}$ & $Q_{XD}$ & $t_{H} \cdot 10^{6}$ & $t_{He} \cdot 10^{6}$
&
$t_{He} / t_{H}$ & CASE\\ 
        &      &        &      &       &      &      &       &
&       &      &        &       &       &\\      
\hline
0.0001	&1     &4.58	& 4.51 & 4.98  &4.41  &  --  & --    & 0.094
&0.300  &0.524 & 8.179  & 1.139 & 0.139 &B\\ 
\hline
0.009	&0.001 &4.58	& 4.51 & 4.98  &4.41  &  --  & --    & 0.163
&0.407  & 0.518 & 8.975  & 0.920 & 0.102 &B\\
	&1   &	4.58	& 4.51 & 4.98  &4.41  &  --  & --    & 0.163  &
	0.407
	& 0.518 & 8.975  & 1.171 & 0.130 &B\\
	&4   &	4.58	& 4.51 & 4.98  &4.41  &  --  & --    & 0.163  &
	0.407
	& 0.518 & 8.975  & 1.279 & 0.142 &B\\
	&5   &	4.58	& 4.51 & 4.98  &4.41  &5.12  & 3.87  & 0.163  &
	0.407
	& 0.518 & 8.975  & 1.155 & 0.128 &A\\
	&100 &	4.58	& 4.51 & 4.98  &4.41  &5.12  & 4.08  & 0.163  &
	0.407
	& 0.518 & 8.975  & 0.851 & 0.095 &A\\
	&500 &	4.58	& 4.51 & 4.98  &4.41  &5.12  & 3.89  & 0.163  &
	0.407
	& 0.518 & 8.975  & 0.915 & 0.112 &A\\ 
	&800 &	4.58	& 4.51 & 4.98  &4.41  &--    & --    & 0.163  &
	0.407
	& 0.518 & 8.975  & 1.050 & 0.117 &C\\ 
\hline
0.015	&5  &	4.58	& 4.52 & 5.02  &4.39  &  --  & --    & 0.206
&
0.447  & 0.541 & 9.520  & 0.854 & 0.089 &C\\
	&40  &	4.58	& 4.52 & 5.02  &4.39  &  --  & --    & 0.206  &
	0.447
	& 0.541 & 9.520  & 0.867 & 0.091 &C\\
	&45  &	4.58	& 4.52 & 5.02  &4.39  &5.17  & 4.25  & 0.206  &
	0.447
	& 0.541 & 9.520  & 0.847 & 0.089 &A\\
	&100 &	4.58	& 4.52 & 5.02  &4.39  &5.17  & 4.26  & 0.206  &
	0.447
	& 0.541 & 9.520  & 0.995 & 0.104 &A\\ 
	&120 &	4.58	& 4.52 & 5.02  &4.39  &  --  & --    & 0.206  &
	0.447
	& 0.541 & 9.520  & 0.835 & 0.087 &C\\ 
        &200 &	4.58	& 4.52 & 5.02  &4.39  &  --  & --    & 0.206
        &
        0.447  & 0.541 & 9.520  & 0.817 & 0.085 &C\\ 
\hline
0.018	&20  &	4.58	& 4.52 & 5.04  &4.38  &--    & --    & 0.238
&
0.464  & 0.562 & 9.805  & 0.783 & 0.079 &C\\
	&50  &	4.58	& 4.52 & 5.04  &4.38  &5.18  & 4.12  & 0.238  &
	0.464
	& 0.562 & 9.805  & 0.805 & 0.082 &A\\
	&100 &	4.58	& 4.52 & 5.04  &4.38  &5.18  & 4.16  & 0.238  &
	0.464
	& 0.562 & 9.805  & 0.837 & 0.085 &A\\
	&280 &	4.58	& 4.52 & 5.04  &4.38  &5.18  & 4.06  & 0.238  &
	0.464
	& 0.562 & 9.805  & 0.815 & 0.083 &A\\
        &300 &	4.58	& 4.52 & 5.04  &4.38  &--    & --    & 0.238
        &
        0.464  & 0.562 & 9.805  & 0.845 & 0.086 &C\\ 
        &400 &	4.58	& 4.52 & 5.04  &4.38  &--    & --    & 0.238
        &
        0.464  & 0.562 & 9.805  & 0.762 & 0.077 &C\\ 
\hline
0.021	&0.1 &	4.58	& 4.51 & 5.06  &4.36  &--    & --    & 0.264
&
0.486  & 0.587 &10.060  & 0.872 & 0.086 &C\\ 
	&1   &	4.58	& 4.51 & 5.06  &4.36  &5.20  & 4.11  & 0.264  &
	0.486
	& 0.587 &10.060  & 0.839 & 0.083 &A\\ 
	&5   &	4.58	& 4.51 & 5.06  &4.36  &5.20  & 4.19  & 0.264  &
	0.486
	& 0.587 &10.060  & 0.983 & 0.097 &A\\ 
	&50  &	4.58	& 4.51 & 5.06  &4.36  &5.20  & 4.21  & 0.264  &
	0.486
	& 0.587 &10.060  & 0.759 & 0.075 &A\\ 
	&300 &	4.58	& 4.51 & 5.06  &4.36  &5.19  & 4.05  & 0.264  &
	0.486
	& 0.587 &10.060  & 0.746 & 0.074 &A\\  
	&500 &	4.58	& 4.51 & 5.06  &4.36  &--    & --    & 0.264  &
	0.486
	& 0.587 &10.060  & 0.746 & 0.074 &C\\  
\hline
0.027   &50  &  4.58    & 4.51 & 5.11  &4.33  &--    & --    & 0.310
&
0.528  & 0.633 &10.573  & 0.745 & 0.070 &C\\ 
\hline
0.05    &100 &  4.58    & 4.51 & 5.21  &4.30  &--    & --    & 0.415
&
0.659  & 0.755 &13.571  & --    & --    &-\\ 
\hline   
\end{tabular*}
\end{center}
\end{scriptsize}
\end{table*}

\subsection{Mass-loss rates: the canonical prescription }

Throughout the various evolution phases up to
the so-called de Jager limit 
the mass-loss rates
 are 
from the popular compilation by de Jager et al. (1988), 
although we are aware that Lamers \& Cassinelli (1996) have recently 
emphasized that  the de Jager et al. (1988)  sample is heavily affected by 
selection effects, proposing a new
expression of $\dot M$ for early type stars. Our choice is motivated by
the request of homogeneity with previous work to which the present models are
compared. 

Beyond the de Jager
limit, the mass-loss rate is increased to $\rm 10^{-3} M_{\odot}
yr^{-1}$, as suggested by observational data for the LBV (Maeder \& Conti 
1994). 

As far
as the  dependence of the mass-loss rates on metallicity  is concerned, we
scale the rates according to $(Z/Z_{\odot})^{0.5}$, as
indicated by the theoretical models of stellar wind by Kudritzky et
al. (1989). We adopt this formulation for the sake of comparison
 with
Fagotto et al. (1994) though  the observational data suggest a 
slightly steeper trend, cf. Fig.~1 in Lamers \& Cassinelli
(1996). Note also that theoretical mass-loss
rates for the most massive stars could be down by a significant factor
with respect to the real ones (de Koter et al. 1997).

The WNL stage of WR stars is assumed to start when the surface abundance of
hydrogen falls below X=0.40, as in Maeder (1990). In
this phase we adopt $\dot M = 10^{-4.4} {\rm M_{\odot} yr^{-1} }$. 
During the subsequent  WNE and WC phases, determined by 
the condition on the abundances X $<$ 0.01
and $\rm C_{sup} > N_{sup}$ (by number) respectively, we apply the 
type-independent relation

\begin{equation}
\dot M = 0.8~10^{-7} (M/M_{\odot})^{2.5}, 
\end{equation}

\noindent
a mean between the values
proposed by Langer (1989) for each sub-group.

The above prescriptions for the
mass-loss rate are of general application and not meant for the
 20 \msol\  star alone. We anticipate however, that the 20 \msol\ 
star always remains
in  the de Jager et al. (1988) regime, and that 
its mass-loss rates during the  RSG phase 
are significantly lower than observed. In
particular, the rates are lower by a factor of 5 as compared to the 
 estimate by Feast (1991) for a sample of RSG stars (see below).

\begin{table*}
\caption{Surface abundances by mass for the models of 20 \msol\
and chemical composition [Z=0.008, Y=0.25]. 
The initial values of the abundances are: He=0.25  C=0.1371e-2
O=0.3851e-2 N=0.4238e-3. }
\begin{scriptsize}
\begin{center}
\begin{tabular*}{120mm}{cc|c|c|c|c|c|c|c|c} 
\hline
$\alpha_{1}$ & $\alpha_{2}$ &$He/He_{i}$    &$C/C_{i}$&  $O/O_{i}$&
$N/N_{i}$&$He/He_{i}$    &$C/C_{i}$&  $O/O_{i}$&
   $N/N_{i}$\\ 
\hline 
\hline
  0.0001& 1    &        1.301    &  0.5446 &     0.8052&
  4.151&-&-&-&-\\ 
\hline
  0.009 & 0.001 &      1.211    &  0.6629 &     0.8611&
  3.275&-&-&-&-\\
  &          1 &         1.3    &   0.581 &     0.8068&
  4.016&-&-&-&-\\
  &          4 &       1.334    &  0.5588 &     0.7868&
  4.259&-&-&-&-\\
  &          5 &       1.309    &  0.5504 &      0.792&       4.257 &
  1.000& 0.9993& 1.000& 1.000\\
  &        100 &       1.168    &  0.672  &      0.878&       3.117&
  1.000& 0.9978&  1.000&  1.000 \\
  &        500 &       1.179    &  0.6377 &     0.8689&       3.325&
  1.024& 0.8344& 0.9771& 1.728\\
  &        800 &       1.207    &   0.608 &     0.8486&
  3.598&-&-&-&-\\ 
\hline
  0.015 &    5 &       1.231    &  0.6446 &     0.8333&
  3.568&-&-&-&-\\
   &         40 &      1.27    &  0.6187 &     0.8063&
   3.893&-&-&-&-\\
   &         45 &      1.335    &   0.423 &     0.7123&       5.352
   &
   1.290& 0.6004& 0.7928& 4.070\\
   &        100 &      1.391    &  0.3535 &     0.6723&       5.911 &
   1.298 & 0.5931& 0.7873& 4.141\\
   &        120 &      1.245    &  0.6424 &     0.8239&
   3.662&-&-&-&-\\ 
   &        200 &      1.265    &  0.6215 &     0.8099 &
   3.853&-&-&-&-\\ 
\hline
 0.018   &  20     & 1.292& 0.6164& 0.7886& 4.037&-&-&-&-\\
         &  50     & 1.356& 0.3799& 0.7003& 5.609 &  1.327& 0.5794&
         0.7627&
         4.387\\
         &  100    & 1.322& 0.5233& 0.7673& 4.519 &  1.314& 0.5964&
         0.7728&
         4.240\\
   &  280    & 1.296& 0.6004& 0.7832& 4.129 &        1.295& 0.6067&
   0.7834&
   4.113\\
   &  300    & 1.274& 0.6232& 0.7993& 3.924&-&-&-&-\\
   &  400    & 1.275& 0.6163& 0.7951& 3.981&-&-&-&-\\ 
\hline
 0.021&      0.1&      1.395    &  0.5602 &     0.7278&
 4.747&-&-&-&-\\
     &        1 &      1.344    &  0.5712 &     0.7525&       4.485 &
     1.334& 0.5921& 0.7543& 4.405\\ 
   &          5 &      1.344    &  0.4958 &     0.7541&       4.693 &
   1.334& 0.5921& 0.7543& 4.405\\  
   &         50 &      1.327    &  0.5196 &     0.7543&       4.618 &
   1.327& 0.5918& 0.7543& 4.398\\
   &        300 &      1.248    &  0.6372 &     0.8016&       3.839 &
   1.248& 0.6372& 0.8019& 3.839\\
   &        500 &      1.248    &  0.6372 &     0.8016&
   3.839&-&-&-&-\\ 
\hline
\end{tabular*}
\end{center}
\end{scriptsize}
\end{table*}

\subsection {Results for the 20 \msol\ star}

As already anticipated, there is a growing amount of evidence
suggesting that a typical 20 \msol\ star with the chemical
composition of the LMC should evolve according to the  evolutionary
scheme of case A after Chiosi \& Summa (1970), which
is characterized by an extended loop across the HRD
during the central He-burning phase. This
is substantiated by observed surface abundances of blue
supergiant stars, suggesting that they have already undergone the
first dredge-up episode in the RSG phase. 

Given these premises, many evolutionary sequences of the 20 \msol\ 
star have been calculated for different choices of the parameters 
$\alpha_1$ and $\alpha_2$ to understand under which circumstances the 
above evolutionary scheme is recovered. The results are reported in 
Table~1. 
Columns (1) and (2) of Table~1 list the parameters $\alpha_1$ and $\alpha_2$
adopted for each sequence. Columns (3) through (8)  
are grouped according to the evolutionary stage; they refer to:
 the zero age main sequence (ZAMS), the stage of minimum effective
temperature during central H-burning (TAMS), the largest extension of
the loops, if present (LOOP). For each stage luminosity and effective
temperature are listed. Columns (9), (10), and (11)
list the following quantities: $ Q_{HE}$,  
the fractionary mass of the He-core at central H exhaustion;
$Q_{XP}$, the fractionary mass at the mid point throughout
the region with a chemical gradient as seen  at
the  core H-exhaustion stage; $Q_{XD}$,  the fractionary mass of the
layer at which  
the H-abundance starts  to vary from surface values  as seen at the
H-exhaustion stage.  Columns (12) and (13) list $t_{H}$ and $t_{He}$,
i.e.  the duration of the
H-burning and He-burning phases, respectively (lifetimes in units of
 $10^{6}$ yr). Column (14) is the He- to H-burning lifetime
ratio. Finally
Column (15) lists the type of each evolutionary sequence according to
the below classification scheme.

\noindent
From the entries of Table 1, three evolutionary regimes are envisaged:

\begin{itemize}
\item Case A: after the main sequence phase the star evolves to the
RSG phase, performs an extended blue loop and completes  He-burning as a
RSG again. This corresponds to the models of Chiosi \& Summa
(1970)
in which no mass loss and the Ledoux neutrality criterion for
semiconvection were adopted.

\item Case B: after the main sequence phase the star begins the
central He-burning phase as a BSG and slowly moves toward the red
side of the HRD, where it ends the He-burning as a RSG. This
corresponds to the models of Chiosi \& Summa  (1970) in
which no mass
loss and the Schwarzschild neutrality condition for semiconvection
were adopted.

\item Case C: in this case  the star spends the whole He-burning
phase as RSG, corresponding to models with large and full
 overshoot. See also Bressan et al. (1993) with
$\Lambda_{c}= 1$ in their formulation.
\end{itemize}

\begin{figure}[h]
\psfig{file=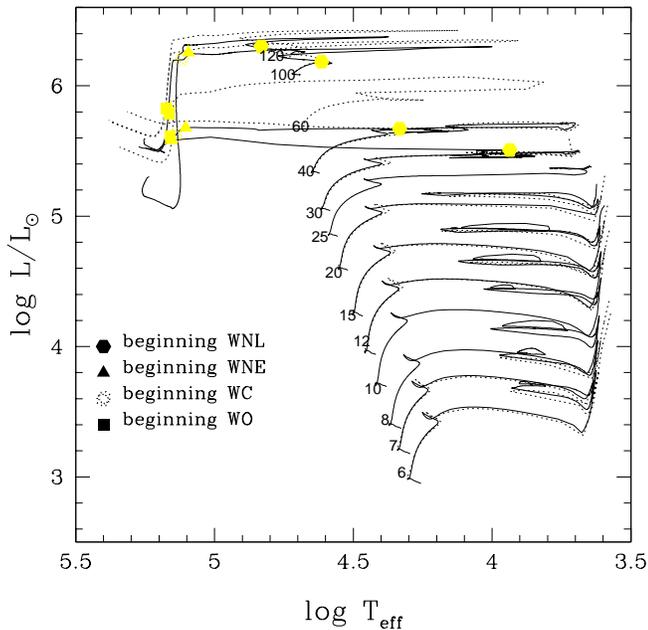,width=8.8cm}
\caption{The set of evolutionary tracks with composition [Z=0.008, Y=0.25]
calculated with the new prescription for diffusive mixing (solid
lines)
compared with the tracks by Fagotto et al. (1994) (dashed lines) with
standard overshoot. The rate of mass loss during the various phase follow
 the standard scheme and are the same for all models (see the text 
for more details).}
\label{traccez008}
\end{figure}

\begin{figure}[h]
\psfig{file=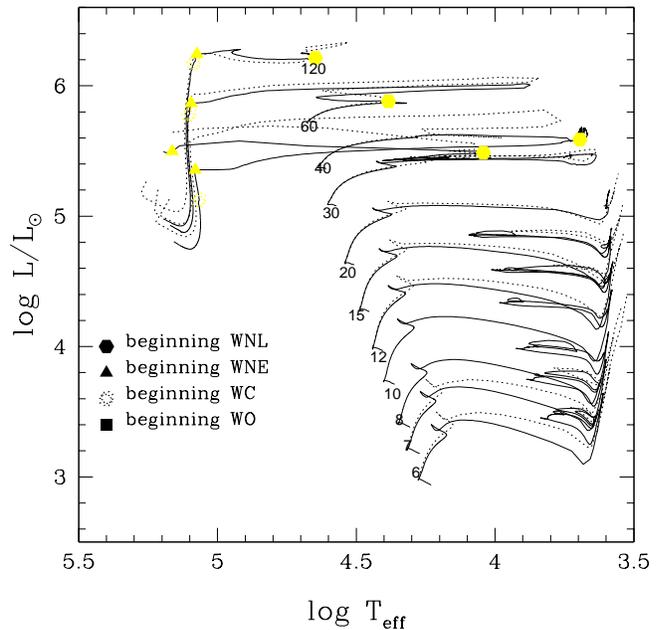,width=8.8cm}
\caption{The same as in Fig.~5 
but for the chemical composition [Z=0.02, Y=0.28].
The new stellar models are compared with those by  Bressan et al. (1993)
with the same composition (dashed lines).}
\label{traccez02}
\end{figure}

The other important feature of the stellar tracks to look at is
the extension of the BHG predicted by the theory.
From the 20 \msol\ star models contained in Table 1,
we can draw the following conclusions:

(1) A large value of $\alpha_1$ yields a  large extension of the
overshoot region and hence a big H-exhausted core ($Q_{HE}$). As long
known this causes a large extension of the main sequence band, a
high luminosity during it and subsequent  phases and, finally, a
long H-burning lifetime. The models presented in Table 1 show
complete mixing in the  overshoot region up to a distance $\approx 13
\alpha_{1} H_P$ from the Schwarzschild border. Therefore they range
from the no overshoot case ($\alpha_{1}=0.0001$) to the extreme
overshoot case ($\alpha_{1}=0.05$), the latter case corresponding to
the models of Bressan et al. (1981).

(2) The size of the core  at the H-exhaustion stage is the factor 
dominating the entire subsequent evolution in the HRD. 
In fact, independently of
 $\alpha_{2}$, i.e.  the mixing efficiency in the
intermediate convective zones, the models with $\alpha_{1} \geq
0.015$ begin the He-burning phase as red supergiants, while the
models with $\alpha_{1} \leq 0.0001$ (no overshoot) do it as blue
supergiants. 

(3) The models  with $\alpha_{1}=0.009$ are of particular  interest
 because they
represent a transition case.  All the three  schemes B, A,
C are indeed  recovered  by increasing the value of  $\alpha_{2}$, i.e. by
decreasing the efficiency of mixing inside the intermediate
convective zone. We note that the fractionary core mass
at central H-burning exhaustion is $Q_{HE}=0.163$,  very close
to the value 0.161 obtained by Deng  (1993) with $P_{diff}=0.4$ in
his formulation.

(4) Although the loop phenomenon eludes simple physical interpretations
as pointed out long ago by  Lauterborn et al. (1971), 
the models we have calculated indicate that when
$0.009 \leq \alpha_{1} \leq 0.021 $, and 
 $50 \leq \alpha_{2} \leq 100$ extended loops are possible.

Many of the results we have obtained  
apparently agree with those  by Langer et al. (1989),
however with major points of difference.
The models by Langer et al. (1989) show indeed a similar
trend in the HRD  at varying his  parameter $\alpha$ (models of
type  B and C are obtained for $\alpha > 0.01$ and  and $\alpha < 0.008$
respectively). However, the 
prescription by  Langer et al.  (1989, 1991) does not favour the occurrence
of convective overshooting. In fact, in presence of 
even modest overshoot (0.15 $H_{P}$ corresponding roughly to
$\alpha_{1}=0.009-0.015$ in our formalism), the models are always of
type C so that they hardly match the observations.

Table 2 adds to the data  of Table 1  the information about 
the surface abundances of the stellar models. We list the ratios of the 
surface abundances  of  $\rm ^{4}He$, $\rm ^{12}C$, $\rm^{16}O$,  
$\rm ^{14}N$ with respect to their initial values. 
The abundances are given at the stage of central
He-exhaustion, Columns (3) through (6), and while the models are in the blue 
loop, Columns (7) through (10).

By inspecting Tables~1 and  2 we suggest that  
$\alpha_1$=0.015 and $\alpha_2$ in the range 50 to 100
should yield stellar models best matching  the general
properties of the HRD of massive stars in the luminosity range
$-7 \geq M_{Bol} \geq  -9$.
 In fact, the blue loops acquire then their maximum extension and  the
 surface abundances of  He and  CNO elements  are
in good agreement with 
observations, eg. Fitzpatrik $\&$ Bohanna (1993). For  
slightly lower  values of $\alpha_1$ the blue loops shrink and, 
more important, the surface
abundance of He and CNO elements are only in very marginal agreement with 
 observations.
The situation gets worst at increasing $\alpha_1$.  
Finally, values of $\alpha_1 > 0.021 $ can certainly be excluded as 
they always lead to C-type models.

The case  $\alpha_1$=0.015 corresponds to a completely mixed
overshoot region of about 0.2 $H_P$ above the Schwarzschild border.
This value is significantly smaller than what is suggested by the
hydrodynamical models of Grossmann and  Xiong.  Had we adopted
$\alpha_1 \simeq$0.5 as indicated  by the former studies, we would have
gotten only the case C evolution, 
in disagreement with the observed distribution of stars across the  HRD.  
It is worth noticing that Bressan et al. (1993),
Fagotto et al. (1994) and the Geneva 
group (Charbonnel et al. 1993, Schaerer et al. 1993) adopt 0.25  H$_P$ above the
Schwarzschild border, a value that corresponds to about
$\alpha_1$=0.018. This explains why the extension of the blue loops of
those models was always too short or missing at all.

Values of $\alpha_2$ for which  extended blue loops do occur are
generally much larger than 1. This indicates that the characteristic
time of the mixing process is larger than the thermal dissipation
time-scale or equivalently  the growing time of the oscillatory convection. 
Full mixing inside the intermediate unstable region, i.e. the straight
application of the Schwarzschild criterion, is excluded by the present
computations. However, no mixing at all  can
be also excluded, because the corresponding models would behave 
as in case C.  Therefore, we are
inclined to conclude  that the characteristic time of the mixing process 
in this region is
slightly longer than the Kelvin-Helmoltz lifetime. In other words,
after central H-exhaustion, by the time  perturbations grow
in the intermediate unstable region the star has already evolved
into the RSG phase.

Finally, our calculations show that the occurrence of the  BHG is a 
general feature of  all these evolutionary
sequences. In fact,  a BHG   about  $\Delta$ Log T$_{eff}$=0.1 wide  
is always predicted to exist between the
maximum extension of the main sequence band and the hottest point of
the He-burning phase. Because the models in Table 1 span  quite  a large range of
mixing efficiencies both in the overshoot and intermediate convective region,
we are convinced that  an important physical ingredient is either 
still missing or badly evaluated in the stellar models   of massive stars.
 We will come back  to this problem  later.

\begin{figure}[h]
\psfig{file=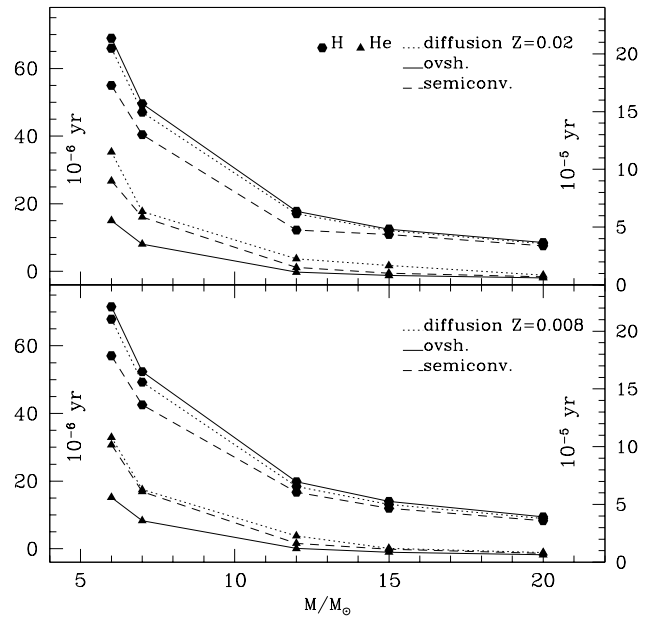,width=8.8cm}
\caption {Comparison of the lifetimes of the  core H- and
He-burning phases for the models with the new diffusive scheme (dotted lines)
 classical semiconvection (dashed lines) by Bressan et
al. (1993), and standard overshooting (solid lines)  by
Fagotto et al. (1994). The top panel is for the chemical composition
[Z=0.020, Y=0.28], whereas the bottom panel is for [Z=0.008, Y=0.25].
The H-burning lifetimes (full dots) are in
units of $10^{6}$ yr (left vertical axis). The He-burning lifetimes
(full triangles) are in units of $10^{5}$ yr (right vertical axis)}
\label{qhel}
\end{figure}

\subsection{ The whole sets of stellar models}

Adopting the parameters $\alpha_1$=0.015 and $\alpha_2$=50 
indicated by the previous analysis of the  20 \msol\  star, 
we have computed two sets of
evolutionary tracks with initial masses 6, 7, 8, 10, 12, 15, 20, 30,
40, 60, 100, 120 \msol\ and chemical composition [Z=0.008, Y=0.25] 
and
[Z=0.020, Y=0.28].  Extensive tabulations of the main physical quantities
for all these stellar models  are not given here 
for the sake of brevity. They are  available from the authors upon request.

The evolutionary path (solid lines) of these stellar 
models on the HRD are shown in Figs.~\ref{traccez008} and \ref{traccez02}
for  Z=0.008 and Z=0.02, respectively.   They are
compared with the corresponding models computed with mild overshoot
(dotted lines) by Fagotto et al. (1994)  and Bressan et al. (1993).

The main sequence band of the new
models is slightly narrower  than in the case of the older tracks.
This  follows from the adopted value of  $\alpha_1$
which corresponds to a smaller overshoot distance. 

Although the fully
mixed region in the models with diffusion is  smaller  than in the 
old ones with  straight homogenization, the amount of fuel
brought into the region of nuclear reactions combined with a slightly
lower luminosity produce an almost equal H-burning lifetime. This is shown by
Fig. \ref{qhel} where the lifetimes $t_H$ and $t_{He}$ for diffusive, 
classical semiconvective, 
and straight overshoot models are plotted as a 
function of the initial mass of the stars 
(top panel for Z=0.02, bottom panel for Z=0.008). 
This fact suggests that part
of the inner chemical profile is due to diffusion itself and 
not to the receding of the convective core as the evolution
proceeds. 

The smaller H-exhausted cores of the new models yield a lower luminosity 
and in turn a longer  core He-burning lifetime. Note in Fig. \ref{qhel} that,
while the H-burning lifetime of the models with diffusion is almost equal
to that of models  with straight overshoot and longer than that of models
with semiconvection alone, the trend is reversed for the He-burning
phase.

Without performing a detailed comparison of the new models with the
observational properties of the HRD in the LMC and  Solar
Vicinity, which is beyond the scope of the present study, yet a number 
of conclusions are possible by simply examining the evolutionary tracks.\\

(1) The BHG  is narrow  in the case of the LMC
metallicity, but it is dramatically large in the case of  solar
metallicity. With the adopted parameters, the 20 \msol\ star with 
Z=0.02 follows case C evolution, which is at  odd with
the stellar census across the HRD of the Solar Vicinity.\\

\begin{figure}[h]
\psfig{file=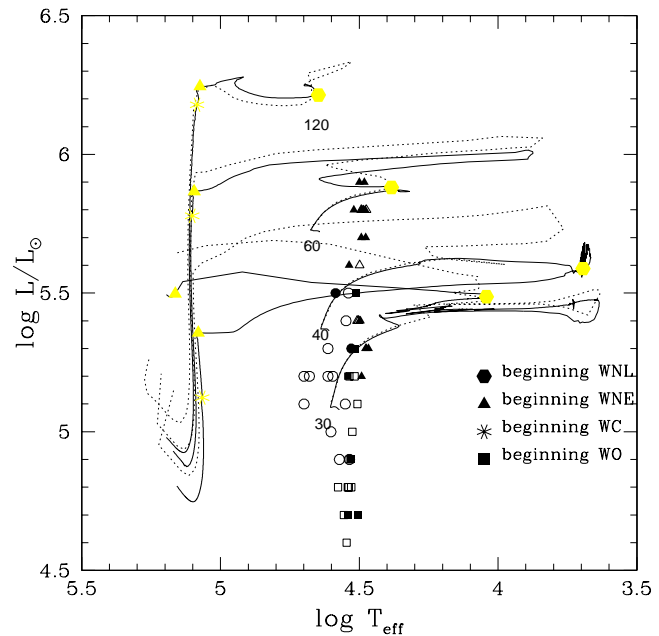,width=8.8cm}
\caption {Comparison of the evolutionary tracks with composition 
[Z=0.02, Y=0.28] during their   WR stages and the observed WNL and WNE stars by
Hamann et al. (1993). The data (triangles, circles and squares) refer
to WNL, $\rm WNE_{s}$ and  $\rm WNE_{w}$ stars,
respectively. Full symbols are WR stars with detected  hydrogen,
whereas empty symbols stand for WR stars with no hydrogen.}
\label{traccez02hamann}
\end{figure}

(2) As long known, the evolution of the most massive stars (M
$\rm \geq 30\div 40$ \msol) is dominated by mass-loss rather 
than internal
convection (cf. Chiosi \& Maeder 1986; Chiosi et al. 1992). 
In this regard, the present models of  30-120 \msol\ stars are much similar
to the previous ones, in which  similar prescriptions for the
mass-loss rate were adopted. 
Perhaps the most intrigued problem in regard to the overall scenario
of massive star evolution is the existence of WR stars and their
genetic relationships with the remaining population of luminous (massive)
stars.
There is nowadays a general consensus that WR stars are
 the descendants of massive stars in late evolutionary stages.
However, extant theoretical models of WR stars do not fully agree 
with their observational counterparts. In particular, it is hard
to explain the location of low-luminosity WR stars. 
The problem is illustrated in 
Fig.~\ref{traccez02hamann}, which  compares the  present evolutionary 
models  for Z=0.02  with the data for galactic WN stars from Hamann et al. 
(1993).
 Surprisingly, WN stars populate a region which
coincides with the main sequence band in the  luminosity range
 $\rm 4.5 \leq log L/L_\odot \leq 6$. WNL stars, indicated by triangles, 
populate the
bright end of the distribution ($\rm 5.2 \leq log L/L_\odot$), whereas WNE
stars populate the faint end ($\rm log L/L_\odot \leq 5.5$). 
In contrast, extant theoretical models predict
 that WNL stars  should evolve horizontally up to 
$\rm log T_{eff} \simeq 5.1$. Note that, with the adopted mass-loss rates, no
WNL-like model  is fainter  than $\rm log L/L_\odot \simeq 5.5$.
 The problem gets worse for 
WE stars, because they are predicted at $\rm log T_{eff}\geq 5.1$ and 
often brighter than $\rm log L/L_\odot = 5.1$, hence
much hotter  (and brighter) than observed. 

It has been argued that the
 discrepancy in the effective temperature can be cured by applying the
well known correction taking into account departure from hydrostatic 
equilibrium and optical thickness in an expanding atmosphere. In fact,
the photosphere of an expanding dense envelope can be different from 
that of a hydrostatic model (cf. Bertelli et al. 1984 and references therein).
However, only a fraction of the observed WNE  stars
shows evidence of a thick atmosphere (circles in Fig.
\ref{traccez02hamann}), while the majority do not. Therefore, our
hydrostatic models seem to be adequate to the present aims.

In addition, it has been suggested that current theoretical
predictions of the mass-loss rate among the most massive O stars are
down  by a significant factor with respect to observed values (de Koter
et al. 1997). In fact Meynet et al. (1994), adopting a mass-loss 
rate during the whole pre-WNE phase which is twice as much as the rate
provided by the standard de Jager et al. (1988) 
prescription, obtain a better agreement as far as the observed WR/O
ratio is
concerned  (Maeder \& Meynet 1994), and get  a fainter
luminosity during
the WC and WO phases. However they run into the same problem as far
as low luminosity WN stars are concerned.

The same difficulty is encountered  even by  more sophisticated models,
which combine the interior structure with a realistic expanding
atmosphere. The reader is referred to   Fig.~8 and related discussion in
Schaerer (1996) for more details. 

Finally,   the ratio between the number of massive stars still on the
main sequence and the  number of WR stars observed in galaxies of the
 Local Group is $\sim$ 3, and  among other things independent of 
metallicity (Massey et al. 1995; Massey et al. 1995a; Massey 1997).
In contrast, the
nuclear time-scales involved suggest a ratio of  $\sim $10,  thus indicating
that under a normal initial mass function the lowest mass for  WR
progenitors   is  smaller  than about  40 \msol\ (cf.
Table~14 in Massey et al. 1995) for the field stars in
LMC. A cautionary remark is worth being made here as the   number ratios 
above could change in the
light of the recent re-classification of WR stars in R136a 
(de Koter et al. 1997).\\

The  results of this section can be summarized as follows
\begin{itemize}
\item{Even these new stellar models cannot reproduce
the observed distribution of O-B-A stars in the HRD in the luminosity
interval $\rm  -7 \leq M_{Bol} \leq -9$ 
(the BHG problem) and the distribution of low luminosity WN stars.} 

\item{The many models  computed with different  mixing
efficiency in  their  convectively unstable  zones  indicate that 
mixing  alone  cannot  solve the problem. }
 
\item{Is mass loss the great villain of the whole story? And could higher 
rates of mass loss  expose inner layers to the surface
even in the mass range $\rm 15 \div 30$ \msol?}
\end{itemize}

Recalling past work along this vein, 
Bertelli et
al. (1984) were able to produce evolutionary 
tracks  of a 20 \msol\
model with the suitable low surface hydrogen abundance, corresponding
to a WN star, through the combined effect of an enhancement in the opacity
around one million $\rm K^o$ and a constant mass-loss rate of a few 10$^{-5}$
\msol/yr during the RSG phase. 

More recently Bressan (1994)  suggested that a larger mass
loss rate in the RSG phase and a suitable treatment of internal
mixing could, at least for the model of 20 \msol\ and solar abundance,
lead  a star to abandon the RSG region and display the features of a
WN star. He noticed that strong support to this idea comes
from the  mass-loss rates for  RSG stars reported by Feast (1991).

\section{ New mass-loss rates in the RSG phase }

The considerations made in the previous section suggest a
re-examination of  the mass-loss rates adopted during the red
supergiant phase  for stars in the mass range 10 to 30 \msol.

It is known (Chiosi \& Maeder 1986; de Koter et al. 1997) that mass-loss
rates in the upper end of the HRD are uncertain by a large factor.  
This is  particularly true  for red supergiant stars, where 
the mass-loss rate is
estimated from a semi-empirical relation obtained by Jura (1986)
for the AGB stars. Uncertainties in the above relation are due to
 the distance of the objects  and 
the gas to dust ratio adopted to convert the dust mass-loss
rate into the total mass-loss rate.

The  average value of the  mass-loss rate  of RSG  stars of  the 
LMC 
is $\rm 3.6 \cdot 10^{-5} M_{\odot}/yr$, whereas the value
derived from the de Jager et al. (1988)  formulation for a typical
20 \msol\ model with  $\rm log T_{eff}$=3.6 and $\rm log
L/L_{\odot}$=5.0 is to $\rm \dot M \sim 10^{-6} M_{\odot}/yr$. 

Recently, Feast (1991) found  a tight correlation between
the mass-loss rate and the pulsational period in RSG stars of the
LMC:

\begin{equation}
log (\dot M) = 1.32\times logP -8.17
\label{feast1}
\end{equation}

\noindent
where $P$ is in days.

A similar empirical relation has been proposed by Vassiliadis and
Wood  (1993)  
for Mira and OH/IR stars, and has been applied to the
evolution of AGB stars. This relation breaks down
when the star reaches a pulsational period of about 500 days;
afterward the mass-loss rate increases at a much slower rate or even remains 
constant with the period. Remarkably in this second stage, the so called
super-wind phase, the mass-loss rate reaches about  the value one
would obtain by equating
the gas-momentum flux to  radiation-momentum flux. 

On the theoretical side, hydrodynamical models by  
Bowen \& Willson (1991) and Willson et al. (1995) 
show that  shock waves generated by  the large amplitude
pulsations in AGB stars levitate matter out to a radius where dust
grains can condensate; after this stage, radiation pressure on grains
and subsequent energy re-distribution by collisions accelerate the
matter beyond the escape velocity.  

A thorough discussion of the problem, in particular whether the ratio
between the two fluxes  may be of the order of unity, 
 can be found in Ivezic \& Elitzur (1995). 

Assuming the distance modulus to the LMC (m-M)=18.5, and
combining eq. (\ref{feast1}) with the following  empirical relation between
the bolometric magnitude and the pulsational period  (Feast 1991)

\begin{equation}
M_{bol}=-2.38 \times logP - 1.46
\label{feast2}
\end{equation}

\noindent
we obtain  the  relation between the mass-loss rate and
the luminosity of the star:

\begin{equation}
log (\dot M)=-8.17 +0.554 \times [2.5 \times log(\frac{L}{L_{\odot}})-6.18]
\label{feast}
\end{equation}

\begin{figure}
\psfig{file=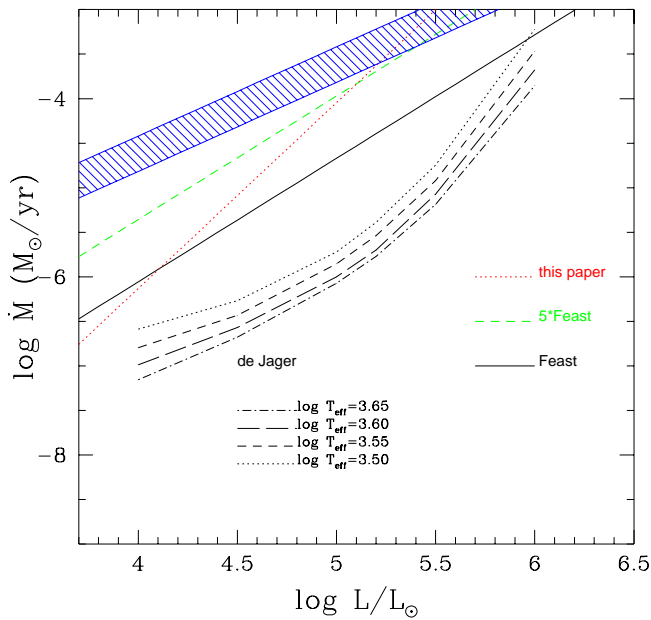,width=8.8cm} 
\caption{Comparison of the mass-loss rates by 
de Jager et al.(1988) (dashed lines) for  different  values of 
$\rm log T_{eff}$ with those by  Feast (1991), solid line. 
 The hatched area corresponds to
the super-wind phase for  the metallicity 0.008$\leq$Z$\leq$0.02 according to
eqs. (\ref{supv})
and (\ref{bet}) in the text. 
The dashed line is the Feast (1991) rate multiplied by a factor of five. 
Finally, the  dotted line shows the prescription for the mass-loss 
rate during the RSG stages we have adopted (see text for all details). }
\label{cfrmloss}
\end{figure}

Relation (\ref{feast}), thereinafter referred to as the {\it Feast
 Relation},  is shown in Fig.~\ref{cfrmloss} (solid line)
together with  the mass-loss rate obtained with de Jager et al.
(1988) (dashed lines) for different effective temperatures,
and the mass-loss rates  for 
the super-wind phase (dashed areas):

\begin{equation}
\dot M = 6.07023 \times 10^{-3}~\beta~ \frac{L}{c v_{\rm
exp}}
\label{supv}
\end{equation}

\noindent
where the mass-loss rate is in \msol\ yr$^{-1}$, the luminosity $L$ is in
solar units, the expansion velocity is v$_{\rm exp}$ = 15 Km s$^{-1}$
and c is the light speed in Km s$^{-1}$. 
$\beta$ is a quantity of order of unity,  in which, 
following the empirical calibration by Bressan et al. (1997),
we include the metallicity dependence 

\begin{equation}
\beta=1.13\times  \frac{Z}{0.008}
\label{bet}
\end{equation}

Note that in the luminosity range characteristic of RSG stars 
($\rm logL/L_{\odot} \approx 4\div 5.4$) 
the mass-loss rate computed according
to Feast (1991),  eq. (\ref{feast}),  is about an order of magnitude 
larger than the one obtained with the de Jager et al. (1988)
formulation. This difference  gives an idea of
the uncertainty with which the rates are   presently known.
Tracing back the causes of this discrepancy is beyond the scope of
the present paper. However we notice that, while the Feast (1991)
formulation  ultimately depends  on the luminosity 
of the star, the relation by de
Jager et al. (1988)  contains  also the effective temperature. If for
any  reason the models fail to reproduce the correct effective
temperature of RSG stars (indeed the models are bluer than the real
stars, a long known  problem), the empirical fit by de Jager et al. 
(1988) does not provide
the right values of the mass-loss rates. 
Furthermore, owing to the very large range of stellar parameters 
encompassed by the de Jager et al. (1998) relations,
a loss of precision in
particular areas of the HRD is always possible. 
Finally, we note that the super-wind mass-loss rate is about 5 times 
larger than the values derived by Feast (1991).
As a conclusion, the current estimates for the mass-loss rates  of
RSG stars  span the range 
$10^{-4}$ to $\rm 10^{-6} M_{\odot}/yr$.

\begin{table*}[t]
\caption{Selected models of the 15 - 18 - 20 \msol\ stars with the new diffusive
scheme and the revised mass-loss rates during the RSG stage. 
Age in years; Masses in solar units; Mass-loss rates in \msol/yr.}
\begin{scriptsize}
\begin{center}
\begin{tabular*}{160mm}{c c c c c c c c c c c c c}
 & & & & & & & & & & & &  \\
\hline
 & & & & & & & & & & & &  \\
 Mod&   Age & $M$ & L/L$_{\odot}$ & T$_{eff}$ & $X_c;Y_c$ & $Q_c$ & $\dot M$  
       & X$_{s}$ & Y$_{s}$ & C$_{s}$ & N$_{s}$ & O$_{s}$ \\
 & & & & & & & & & & & & \\
\hline
 & & & & & & & & & & & &  \\
\multicolumn{13}{c}{20 \msol\ ~~~~ Z=0.020 }\\
 & & & & & & & & & & & & \\
\hline
1& 0.00000E+00 &  20.00 &  4.635 &  4.540 &  0.697 & 0.6337  &  -7.814 & 7.00E-1 & 2.80E-1 & 3.43E-3 & 1.06E-3 & 9.63E-3 \\
2& 7.99360E+06 &  19.29 &  5.011 &  4.375 &  0.020 & 0.3870  &  -6.455 & 7.00E-1 & 2.80E-1 & 3.43E-3 & 1.06E-3 & 9.63E-3 \\
3& 8.13084E+06 &  19.25 &  5.054 &  4.439 &  0.000 & 0.2787  &  -6.461 & 7.00E-1 & 2.80E-1 & 3.43E-3 & 1.06E-3 & 9.63E-3 \\
4& 8.13186E+06 &  19.25 &  5.060 &  4.443 &  0.000 & 0.0847  &  -6.454 & 7.00E-1 & 2.80E-1 & 3.43E-3 & 1.06E-3 & 9.63E-3 \\
 \hline
5& 8.14274E+06 &  19.24 &  5.071 &  4.137 &  0.980 & 0.0000  &  -6.201 & 7.00E-1 & 2.80E-1 & 3.43E-3 & 1.06E-3 & 9.63E-3 \\
6& 8.14865E+06 &  19.22 &  4.890 &  3.587 &  0.979 & 0.1426  &  -4.799 & 7.00E-1 & 2.80E-1 & 3.43E-3 & 1.06E-3 & 9.63E-3 \\
7& 8.15614E+06 &  19.01 &  5.108 &  3.550 &  0.974 & 0.1698  &  -4.504 & 6.41E-1 & 3.39E-1 & 2.18E-3 & 3.81E-3 & 8.01E-3 \\
8& 8.86389E+06 &   8.20 &  5.112 &  4.423 &  0.168 & 0.6753  &  -6.312 & 4.33E-1 & 5.47E-1 & 3.02E-4 & 1.03E-2 & 3.11E-3 \\
9& 9.02589E+06 &   8.10 &  5.221 &  3.732 &  0.000 & 0.4526  &  -5.837 & 4.30E-1 & 5.51E-1 & 2.94E-4 & 1.04E-2 & 3.05E-3 \\
\hline
 & & & & & & & & & & & &  \\
\multicolumn{13}{c}{18 \msol\ ~~~ ~ Z=0.020 }\\
 & & & & & & & & & & & &  \\
\hline
1 & 0.00000E+00 &  18.00 &  4.510 &  4.521 &  0.695 & 0.6090  &  -8.142 & 7.00E-1 & 2.80E-1 & 3.43E-3 & 1.06E-3 & 9.63E-3 \\
2 & 9.08043E+06 &  17.56 &  4.902 &  4.368 &  0.022 & 0.3709  &  -6.674 & 7.00E-1 & 2.80E-1 & 3.43E-3 & 1.06E-3 & 9.63E-3 \\
3 & 9.23889E+06 &  17.52 &  4.945 &  4.427 &  0.000 & 0.2916  &  -6.682 & 7.00E-1 & 2.80E-1 & 3.43E-3 & 1.06E-3 & 9.63E-3 \\
4 & 9.24075E+06 &  17.52 &  4.955 &  4.435 &  0.000 & 0.0666  &  -6.665 & 7.00E-1 & 2.80E-1 & 3.43E-3 & 1.06E-3 & 9.63E-3 \\
\hline
5& 9.25359E+06 &  17.52 &  4.960 &  4.120 &  0.980 & 0.0000  &  -6.384 & 7.00E-1 & 2.80E-1 & 3.43E-3 & 1.06E-3 & 9.63E-3 \\
6& 9.25939E+06 &  17.51 &  4.746 &  3.621 &  0.979 & 0.1192  &  -5.001 & 7.00E-1 & 2.80E-1 & 3.43E-3 & 1.06E-3 & 9.63E-3 \\
7& 9.26727E+06 &  17.34 &  5.019 &  3.584 &  0.974 & 0.1524  &  -4.625 & 6.42E-1 & 3.38E-1 & 2.16E-3 & 3.80E-3 & 8.04E-3 \\
8& 1.02323E+07 &   7.01 &  5.007 &  4.539 &  0.075 & 0.6544  &  -6.824 & 4.32E-1 & 5.49E-1 & 2.46E-4 & 1.04E-2 & 3.06E-3 \\
9& 1.03189E+07 &   6.99 &  5.103 &  3.796 &  0.000 & 0.3468  &  -6.083 & 4.31E-1 & 5.49E-1 & 2.45E-4 & 1.04E-2 & 3.05E-3 \\
\hline
 & & & & & & & & & & & &  \\
\multicolumn{13}{c}{20 \msol\ ~~~ ~ Z=0.008 }\\
 & & & & & & & & & & & & \\
\hline
1 & 0.00000E+00 &  20.00 &  4.607 &  4.552 &  0.725 & 0.6423  &  -8.124 & 7.42E-1 & 2.50E-1 & 1.37E-3 & 4.24E-4 & 3.85E-3 \\
2& 9.19228E+06 &  19.49 &  5.055 &  4.379 &  0.011 & 0.4004  &  -6.568 & 7.42E-1 & 2.50E-1 & 1.37E-3 & 4.24E-4 & 3.85E-3 \\
3& 9.28451E+06 &  19.47 &  5.092 &  4.435 &  0.000 & 0.3288  &  -6.571 & 7.42E-1 & 2.50E-1 & 1.37E-3 & 4.24E-4 & 3.85E-3 \\
4& 9.28656E+06 &  19.47 &  5.098 &  4.441 &  0.000 & 0.0909  &  -6.567 & 7.42E-1 & 2.50E-1 & 1.37E-3 & 4.24E-4 & 3.85E-3 \\
 \hline
5& 9.29425E+06 &  19.46 &  5.110 &  4.233 &  0.992 & 0.0000  &  -6.354 & 7.42E-1 & 2.50E-1 & 1.37E-3 & 4.24E-4 & 3.85E-3 \\
6& 9.29986E+06 &  19.39 &  5.006 &  3.651 &  0.990 & 0.1637  &  -3.983 & 7.42E-1 & 2.50E-1 & 1.37E-3 & 4.24E-4 & 3.85E-3 \\
7& 9.30426E+06 &  18.64 &  5.191 &  3.618 &  0.986 & 0.2019  &  -3.698 & 6.64E-1 & 3.28E-1 & 8.13E-4 & 1.80E-3 & 2.98E-3 \\
8& 9.54444E+06 &   8.82 &  5.031 &  4.392 &  0.619 & 0.6612  &  -6.633 & 6.61E-1 & 3.31E-1 & 8.02E-4 & 1.83E-3 & 2.96E-3 \\
9& 1.00481E+07 &   8.61 &  5.196 &  3.704 &  0.000 & 0.0000  &  -6.043 & 6.22E-1 & 3.71E-1 & 6.09E-4 & 2.42E-3 & 2.55E-3 \\
\hline
 & & & & & & & & & & & &  \\
\multicolumn{13}{c}{15 \msol\ ~~~ ~ Z=0.008 }\\
 & & & & & & & & & & & & \\
\hline
1& 0.00000E+00 &  15.00 &  4.246 &  4.501 &  0.733 & 0.6034  &  -9.189 & 7.42E-1 & 2.50E-1 & 1.37E-3 & 4.24E-4 & 3.85E-3 \\
2& 1.38837E+07 &  14.87 &  4.742 &  4.357 &  0.017 & 0.3549  &  -7.217 & 7.42E-1 & 2.50E-1 & 1.37E-3 & 4.24E-4 & 3.85E-3 \\
3& 1.40510E+07 &  14.86 &  4.787 &  4.413 &  0.000 & 0.2940  &  -7.216 & 7.42E-1 & 2.50E-1 & 1.37E-3 & 4.24E-4 & 3.85E-3 \\
4& 1.40532E+07 &  14.86 &  4.796 &  4.419 &  0.000 & 0.0132  &  -7.203 & 7.42E-1 & 2.50E-1 & 1.37E-3 & 4.24E-4 & 3.85E-3 \\
 \hline
5& 1.40672E+07 &  14.86 &  4.798 &  4.091 &  0.992 & 0.0000  &  -6.857 & 7.42E-1 & 2.50E-1 & 1.37E-3 & 4.24E-4 & 3.85E-3 \\
6& 1.40716E+07 &  14.84 &  4.593 &  3.639 &  0.991 & 0.0696  &  -4.596 & 7.42E-1 & 2.50E-1 & 1.37E-3 & 4.24E-4 & 3.85E-3 \\
7& 1.40783E+07 &  14.22 &  4.977 &  3.608 &  0.987 & 0.1434  &  -3.994 & 6.87E-1 & 3.05E-1 & 8.53E-4 & 1.57E-3 & 3.18E-3 \\
8& 1.45085E+07 &   6.09 &  4.692 &  4.474 &  0.517 & 0.6116  &  -7.609 & 6.85E-1 & 3.07E-1 & 8.45E-4 & 1.60E-3 & 3.16E-3 \\
9& 1.50928E+07 &   6.05 &  4.854 &  3.703 &  0.000 & 0.0000  &  -6.539 & 6.85E-1 & 3.07E-1 & 8.45E-4 & 1.60E-3 & 3.16E-3 \\
\hline
\end{tabular*}
\end{center}
\end{scriptsize}
\end{table*}

\section{ Models with the new mass-loss rate } 
By analogy with the recent models of AGB
stars in which   a relation between the mass-loss rate and the period
has been adopted (Vassiliadis \& Wood 1993), the adoption of  the Feast
(1991) 
formulation would require to test the models for RSG stars 
against 
pulsational instability. On the observational side,
 the majority of galactic RSG show
luminosity variations from a few to several tenths of magnitude
(depending also on the observed spectral region), thus 
suggesting that the RSG phase is most likely pulsationally unstable.
 Heger al. (1997) analyzed the stability of RSG models finding
pulsational instability during the  He-burning phase  of 20 \msol\ and 15
\msol\ stars. A summary of the main properties of the models we are going to
discuss below is given in Table 3, which lists the model number (column 1),
where  
1 through 4  and 5 through 9 correspond to the   core H-burning and He-burning,
respectively;  the  age (column 2), the current
mass (column 3), the logarithm of the luminosity (column 4);
the logarithm of the  effective temperature
(column 5); the central content of hydrogen $\rm X_c$ or helium $\rm Y_c$ 
as appropriate (column 6); the fractionary mass $\rm Q_c$ of the convective 
core (column 7);
the rate of mass loss (column 8), the surface abundance of hydrogen, helium,
carbon,  nitrogen, and oxygen (columns 9 through 13, respectively).

\subsection{The case of solar  composition}

In this section we describe the results  for stars of
 initial masses 12,
18, 20, and 30 \msol\ and composition [Z=0.02, Y=0.28] and
[Z=0.008, Y=0.25] that are calculated using the mass-loss rate of
Feast (1991) given by  eq. (\ref{feast}) during  the RSG stages,
i.e. cooler than  $\rm log T_{eff} < 3.7$. The value adopted for
the boundary effective temperature is somewhat
arbitrary, however, given the dependence of the mass-loss rate on the
sole luminosity and the short time-scale at which this 
area of the HRD is crossed by the models, this choice is less of a problem.

\begin{figure}
\psfig{file=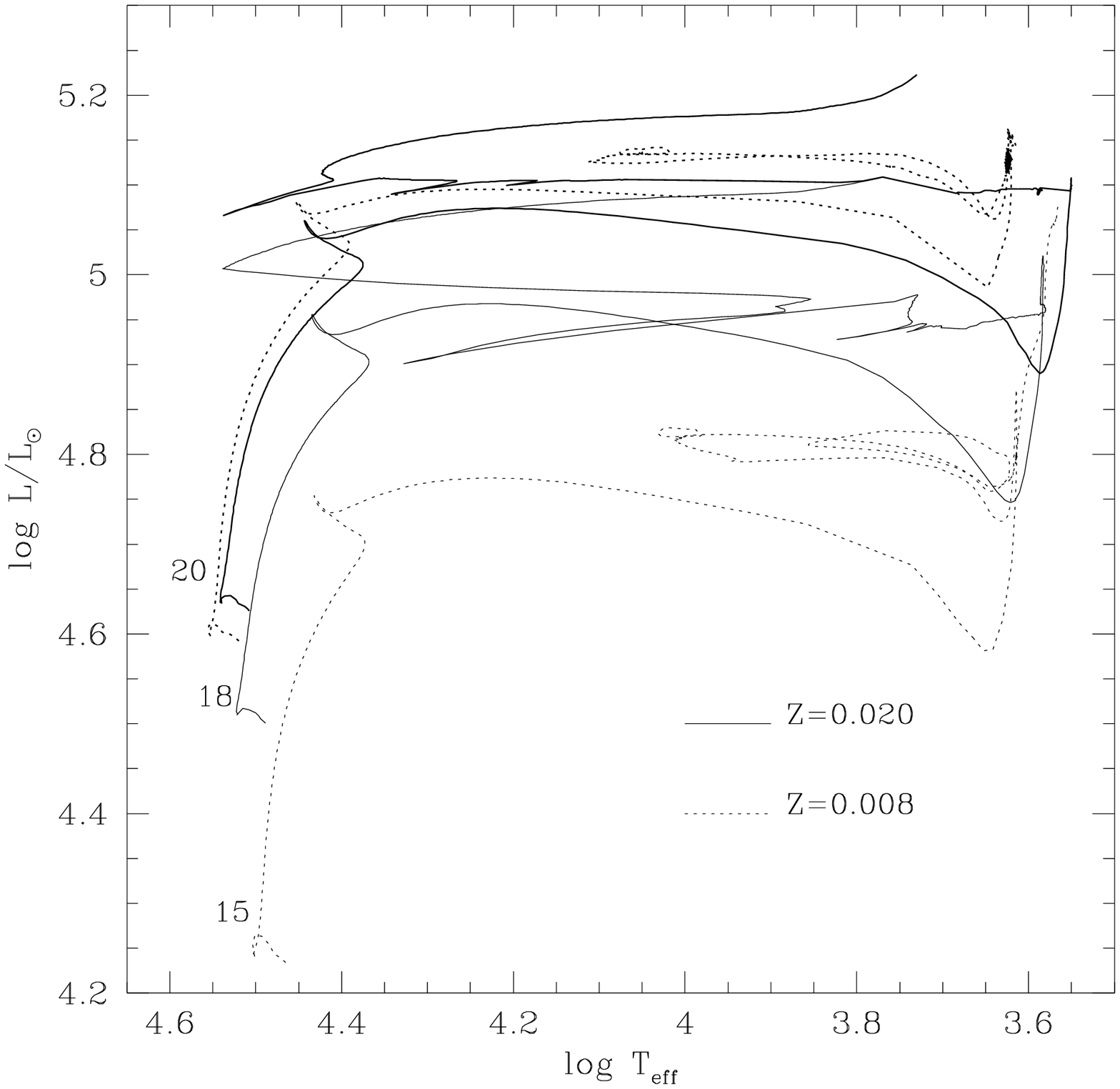,width=8.8cm}
\caption{Evolutionary path in the HRD of 
models calculated with the new diffusive scheme and the mass-loss rate 
during the RSG stages according to Feast (1991). 
The solid lines are for [Z=0.02, Y=0.28],
whereas the dotted lines are for [Z=0.008, Y=0.25]. 
The initial mass is indicated  along  the ZAMS.}
\label{tf1}
\end{figure}

Fig.~\ref{tf1} shows the evolutionary tracks of the  models
computed with the new mass-loss rate in the RSG phase. 

\begin{figure}
\psfig{file=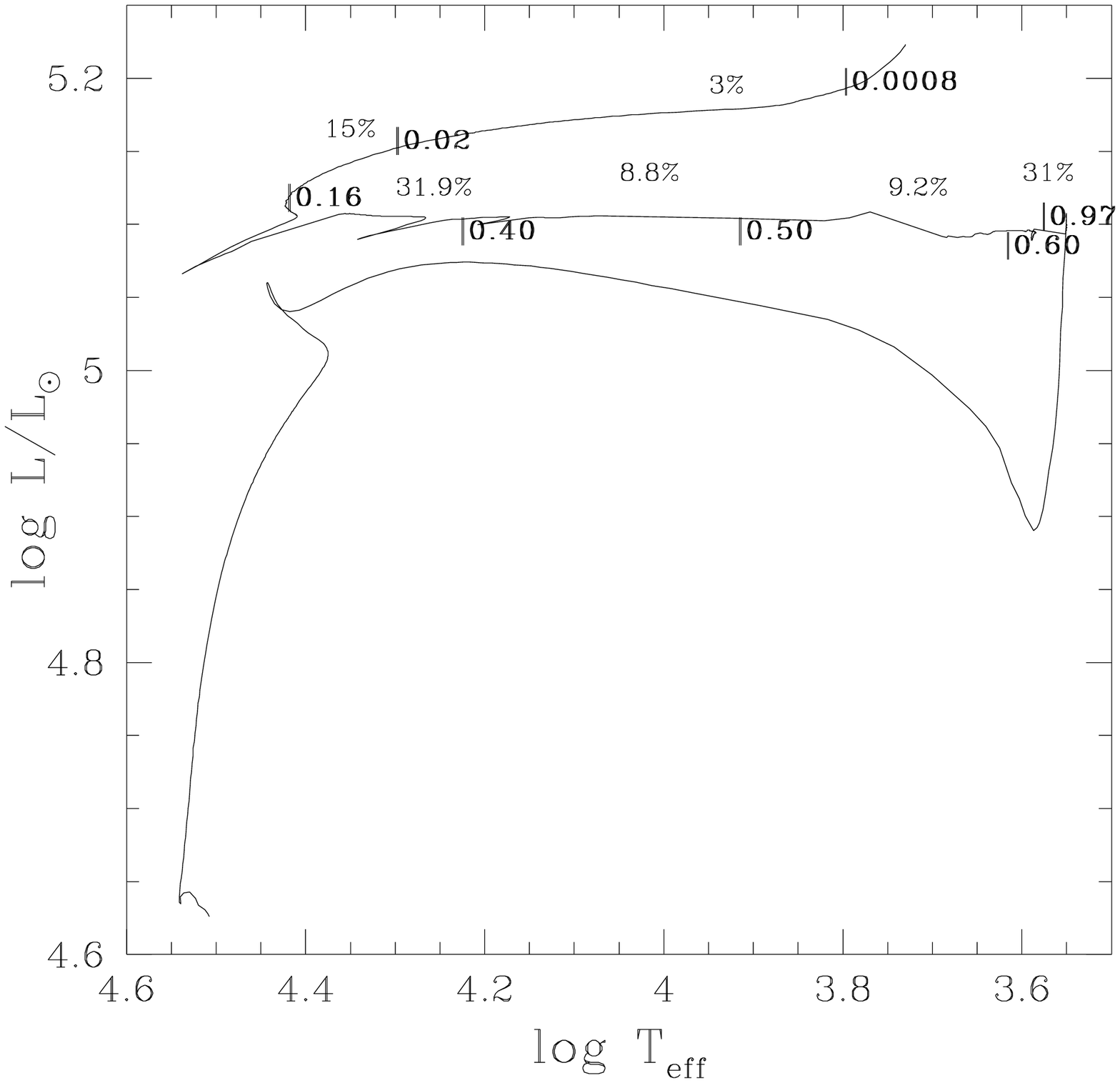,width=8.8cm}
\caption{Path in the HRD of the   20 \msol\ star with
 composition [Z=0.02, Y=0.28], the new diffusive scheme, and the
new mass-loss rate during the RSG phase. Numbers along the tracks indicate 
the central He-abundance and the fractional duration of the He-burning phase
normalized to the whole He-burning lifetime.}
\label{hrm20feast}
\end{figure}

Looking at the  20 \msol\ star as an example, it evolves as in case A
until it reaches the Hayashi line with  total mass of $M
\approx 19$ \msol. The vigorous mass loss ($\simeq
10^{-5}$\msol/yr) in the low effective   temperatures range
 peels off the star leaving a
8.35 \msol\ object which begins a very extended blue loop. The star spends 
$\approx 47 \% $ of the He-lifetime  at $\rm T_{eff}>4.2 $. 
The surface H-abundance gets below X=0.45 already when the central
He-content is $\rm Y_{c}=0.4$ and
$\rm T_{eff}\leq $4.2.  When $\rm Y_{c}$ drops below  $\approx 0.2 $  the
star goes  back to  the red side of the HRD.
The surface H-abundance at the stage of central He-exhaustion is 0.426.
Fig.~ \ref{hrm20feast} displays the
evolutionary path in  the HRD of the 20 \msol\ model in a more
detailed fashion. The  central He-abundance and the fractional duration of
the He-burning phase normalized to the whole He-burning lifetime are
annotated along the evolutionary track.

\begin{figure}
\psfig{file=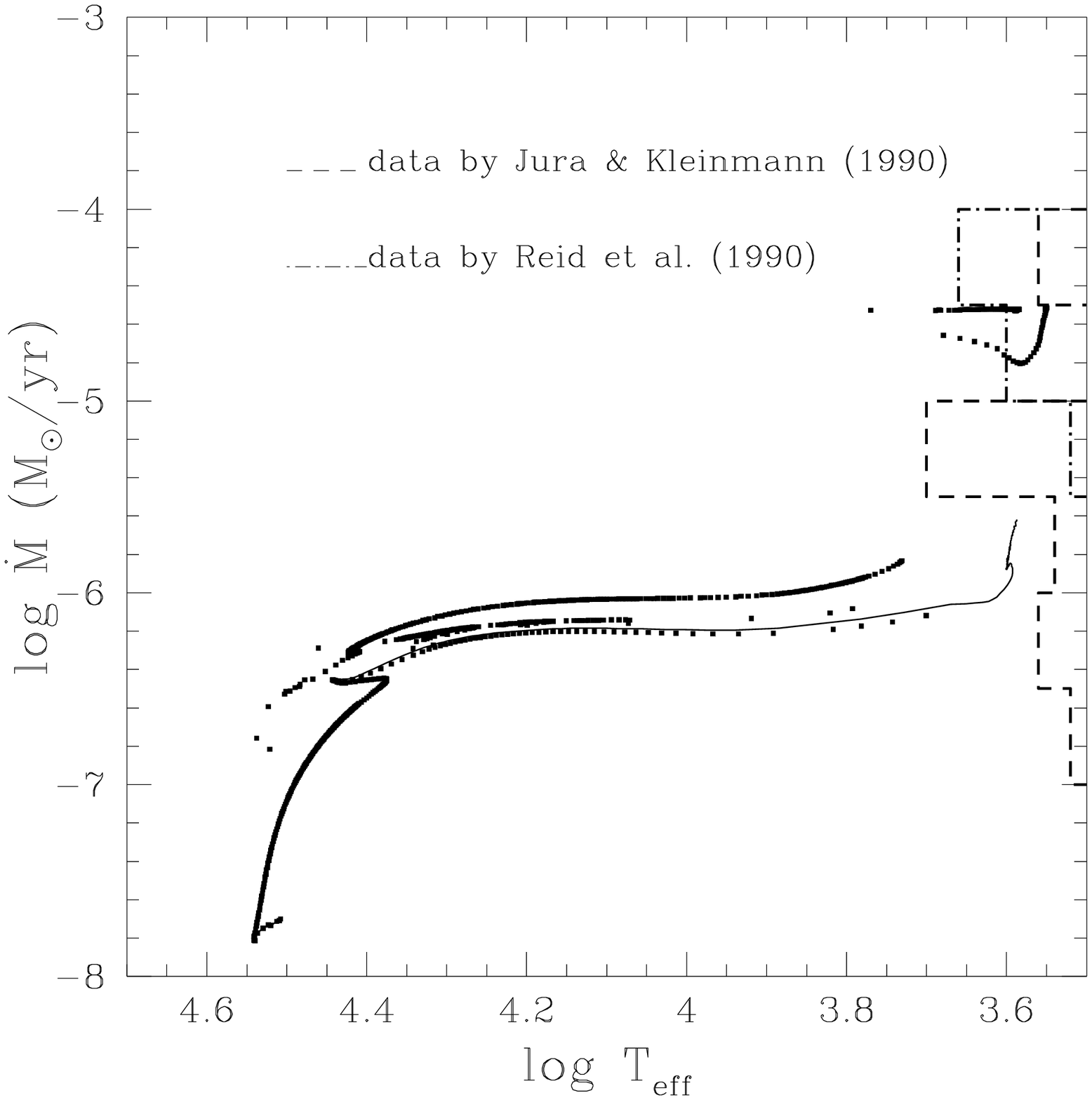,width=8.8cm}
\caption{Mass-loss rate as a function of $\rm log T_{eff}$ for the
20 \msol\ star with composition [Z=0.02, Y=0.28]. The solid
line refers to the standard models with the mass-loss rate by de
Jager et al. (1988). The squares correspond to the models with the
mass-loss rate by Feast (1991) in the RSG stages. 
The histogram on the right vertical axis shows
the observational data by Jura \& Kleinmann (1990) and Reid et al.
(1990).}
\label{isto_mloss}
\end{figure}

As far as the variation of the mass-loss rate in the course of 
evolution is concerned, in Fig.\ref{isto_mloss} we plot  the mass-loss 
rate (in
$\rm M_{\odot}$/yr) as a function of $\rm log T_{eff}$  for the 20 \msol\ star  
calculated with two laws for the mass loss rate, i.e. de Jager et
al. (1988) all over the evolutionary history (solid line) and Feast (1991)
for the   RSG  stages  (points). 
On the right vertical axis of Fig.~\ref{isto_mloss} 
we plot  the histogram of 
the observational data by Jura \& Kleinmann (1990) for 21 red
supergiants within 2.5 kpc of the Sun (dashed line), and by Reid et al.
(1990) for the RSG variables in LMC (dashed-dotted line). 
It is immediately evident that the de Jager et al. (1988) formulation 
severely under-estimates the mass-loss rates in the RSG phase.

The 18 \msol\ stars evolves in a similar fashion, however starting the
blue loop at somewhat later stages, when the central He content is
 $\rm Y_{c}$=0.2 and the mass has been reduced 
to 40\% of its initial mass. 
At the stage of central He-exhaustion,  the
surface H abundance is 0.43 and the total mass is 6.99 \msol. 
A noticeable  difference with respect to the 20
\msol\ star is the smaller duration of the blue loop, $\sim 10\%$ of
$t_{He}$ in this case.

\begin{figure}
\psfig{file=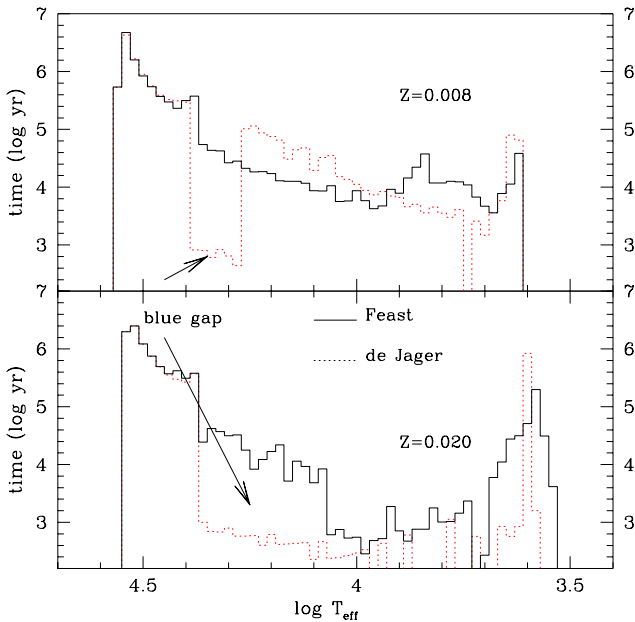,width=8.8cm}
\caption{Histogram of the elapsed time as a function of the
effective temperature for  the 20 \msol\  stars with [Z=0.008, Y=0.250]
 (upper panel) and with [Z=0.020, Y=0.280] (bottom panel). 
The solid lines are for the mass-loss
 rate given by eq.(12), whereas the dotted lines are for the
Jager et al. (1988) prescription.}
\label{istogramma}
\end{figure}

In summary, these models with solar composition and much higher rate of 
mass loss during the RSG phase with respect to the standard  values 
follow case A evolution and possess
a  blue loop that extends into the main sequence band. Clearly these
models are good candidates to solve the long lasting mystery
of the missing BHG. This is shown in the bottom panel of  
Fig.~ \ref{istogramma}, which displays the time (in years on a 
logarithmic scale) spent in various bins of effective temperature across 
the HRD.
The dotted line is for the de Jager et al. (1988) prescription,
whereas the solid line is for the new mass-loss rates in the RSG stages.
It is soon evident that these models would predict a smooth distribution of 
stars across the HRD from the earliest to the  latest spectral types.

\subsection{The case of the LMC  composition}

The above result no longer holds passing to the  metallicity Z=0.008 
typical of the LMC, because even with the new
mass-loss rates the BHG problem is back again.
In fact, the models have shorter blue loops (see Fig.~\ref{tf1}) so that
the main sequence band and the hot side of the blue loops do not merge,
thus producing a gap in between the two.

{\it Can  the problem  be cured by  even higher 
mass-loss rates during the RSG stages? And, in such a case, are the
required mass-loss rates still compatible with the observations? }

To answer the first question we computed, limited to the case of
Z=0.008, a set  of models (15 and 20 \msol) in which the Feast (1991) 
mass-loss rate is
increased by  a factor of 5. Note that, even with this large
increase,  the resulting mass-loss rate is still below the limit imposed
by the super-wind constraints of eqs. (\ref{supv}) and (\ref{bet}).

\begin{figure}
\psfig{file=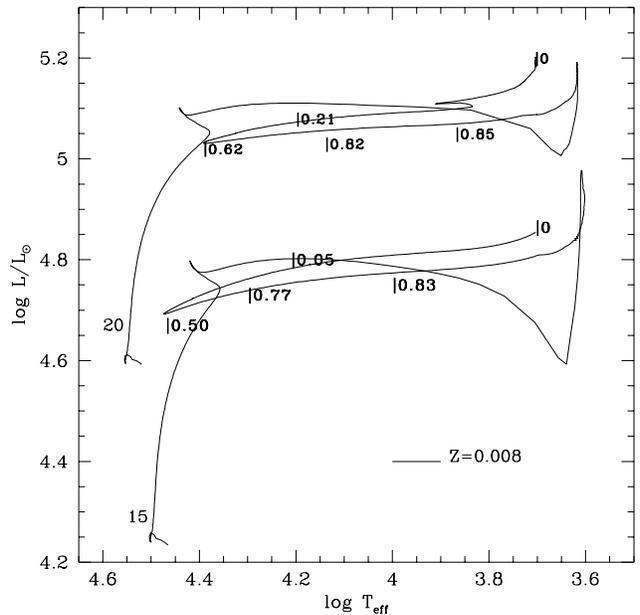,width=8.8cm}
\caption{The path in the HRD of  stellar models calculated
with the  mass-loss rate in the RSG phase by Feast (1991) 
increased by a factor of five. The chemical composition is
[Z=0.008, Y=0.28]. The initial mass is indicated along  the ZAMS, whereas 
the central abundance of helium is annotated along the tracks.}
\label{tf5}
\end{figure}

The corresponding evolutionary tracks are shown in Fig. \ref{tf5}, which
shows once more  that  mass loss  in the  RSG phase is the key
parameter. In fact, the BHG no longer occurs as in the case of  models 
with solar composition.

\noindent
Before answering the second question we note the following: 

\begin{enumerate}
\item{
In the case of solar composition, mass-loss rates larger than adopted 
would not destroy the evolutionary pattern we have found. In fact,  
increasing the mass-loss rate in the RSG phase would
only anticipate the stage at which the models leave the low temperature
 region of the HRD, but
without modifying the structural properties of the models, which 
are ultimately responsible of loop extension toward the blue. }

\item{
In the HRD of Fig.~\ref{tf5}  the extension of the
blue loop is larger at decreasing initial mass. Cast in another way,
this means that the adopted mass-loss rate is larger than required 
at decreasing initial mass. A suitable mass-loss rate should be
about a factor of five larger than the Feast (1991) rate for the 20 \msol\
 model
and only a factor of two for the 15 \msol\ model. 
Not only the
mass-loss rates should be larger than used, but they should also change
their luminosity dependence.} 

\item{
One may argue that we are adopting  mass-loss rates
supposedly standing on pulsational properties without checking a 
priori whether or not our RSG models are pulsationally unstable. 
Secondly we adopt 
mass-loss rates that do not depend on the amplitude of pulsation, which
in contrast if taken into account could increase the dependence on
luminosity, because in AGB stars the amplitude is found to increase with 
luminosity, cf. Heger et al. (1997). 
Finally, current pulsational properties of RSG stars show that they are 
 unstable for a significant fraction of
their lifetime, long enough to lose a significant fraction of their
envelope mass under the proposed mass-loss rates. }
\end{enumerate}

Given these premises,  we revisited the original
determinations of the mass-loss rate for the stars, upon which
the Feast (1991) relation was derived. 

One of the basic assumptions in the
mass-loss estimates made by Jura \& Kleinmann (1990) and  Reid et al. 
(1990) is
the value of the  dust to gas  ratio $\delta$ used  to convert the mass-loss
rate referred to the dust into the total mass-loss rate.
 By analogy with AGB stars,  Jura (1986)  adopted 
 $\delta=0.045$ noticing, however, that a value as low as
0.001 could also be possible.  Since the total rate of mass loss is expected
to  be inversely proportional to $\delta$, the immediate consequence
follows that the total mass-loss rates could be under-estimated by a
a factor of 4.5 (close to our provisional suggestion).

If the high mass-loss rates in the RSG phase are ultimately driven by
the transfer of the photon momentum  to  dust grains and  
gas, a tight  relationship between the dust abundance and the
velocity of the flow is expected. The problem has been studied by
Habing et al. (1994), who confirmed that the terminal velocity  $v_{\rm exp}$ 
of the gas flow   in AGB stars depends rather strongly on
the dust to gas ratio $\delta$. 

Along the same kind of arguments,  
Bressan et al. (1997) obtained the following expression
for $\delta$:

\begin{equation}
\delta \simeq 0.015 \times v_{\rm exp}^2 [{\rm km/s}] \times
            \left( {L\over L_\odot}\right)^{-0.7}
\label{eq_delta}
\end{equation}

Inserting in this relation $v_{exp}=15~ {\rm Km~s^{-1}} $ and
$\rm L=10^4~L_{\odot}$ (typical  of AGB stars) one gets
$\delta$=0.005, which is nearly  the value adopted by
Jura \& Kleinmann (1990) and Reid et al. (1990), thus confirming
 that this is the  value suited to  Mira and OH/IR stars.

In contrast,  if we keep the velocity constant (though for periods above 
500 days there are hints  for a slightly increase   with 
the period) 
 and assume $\rm L=10^5 L_{\odot}$ (typical of RSG star),  we get
$\delta~\simeq$~0.001 indicating that at these high luminosities the
mass-loss rates could have been under-estimated by a significant
factor.

In order to  include the effect   
 of a systematic variation of the dust to gas ratio 
at increasing stellar luminosity on the mass-loss determination, as suggested by 
eq. (\ref{eq_delta}),  
 we multiply the Feast (1991)
relation by the factor 0.0045/$\delta$. The new mass-loss rate is given
by

\begin{equation}
log (\dot M)= 2.1\times log(\frac{L}{L_{\odot}})-14.5
\label{bress}
\end{equation}

This law is shown by the dotted line in Fig.~\ref{cfrmloss}. The
new  relation  is much steeper than the old one by  Feast (1991). 
The mass-loss rate is about 2 times larger
for a 15 \msol\ model and about 5 times larger for a 20 \msol\ model.
The mass loss rate gets the super-wind regime at about
$\rm log(L/L_{\odot}) \simeq 5.2$.

Finally, we like to call attention on  the amazing coincidence between 
the rates  demanded by the properties of
the HRD and those indicated by  our consistent revision of the
observational data. The subject deserves further studies by means of 
the ISO observations of
winds of RSG stars, and  a thorough
investigation of the pulsational properties of RSG stars.

\section{Internal mixing or mass loss ?}

The aim of this section is to understand which between mixing and mass loss 
is the key factor 
determining the kind of models presented above that
apparently succeed to account for the distribution of massive stars
in the HRD.

\begin{figure}
\psfig{file=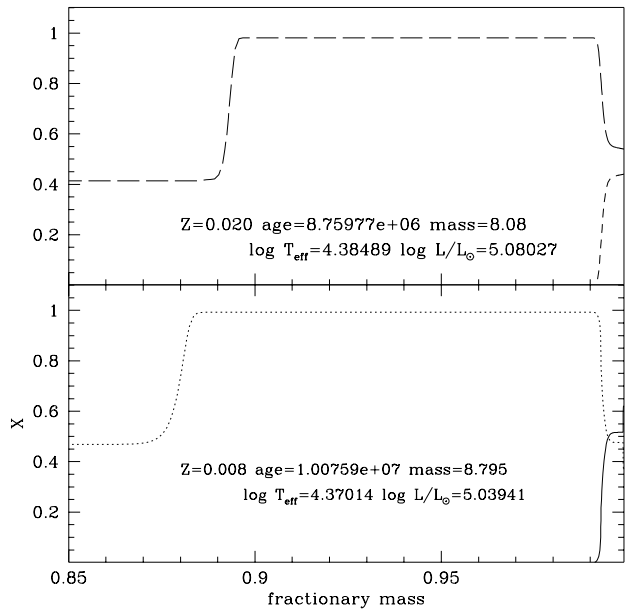,width=8.8cm}
\caption{The internal H- and He-profiles of the 20 \msol\ models with 
[Z=0.020, Y=0.28] (upper panel: dashed and long-dashed lines, respectively)
and [Z=0.008, Y=0.25] 
(bottom panel: continuous and dotted lines, respectively)
during the blue loop. Masses are in solar units.}
\label{prof1}
\end{figure}

The internal structure of all the models during the blue supergiant phase
(Fig. \ref{prof1}) is very similar. The helium core is
surrounded by a thin hydrogen envelope, whose fractionary mass is
about 1\%-5\% of the current total mass. This characteristic seems an
essential condition in order to get core  He-burning models 
in the effective temperature range 
$\rm 4.5 \geq logT_{eff} \geq4.2$. 
Clearly the high  mass-loss rates we have adopted strip  
the models of their envelopes and 
provide a simple explanation of the blue extension of the loop.

We have already seen that internal mixing determines the
evolutionary pattern in
the HRD, i.e. dictates which of the three avenues (case A, B and C) 
is followed by a star.
Furthermore, with the standard mass-loss rates 
(in the RSG stages in particular),
we  determine the values of  $\alpha_1$ and $\alpha_2$ yielding 
the widest loops across the HRD.
In our new evolutionary scenario, 
the role of internal mixing, both in the core and
in the intermediate convective shell,  is less important. 
Nevertheless, case B (slow redward evolution) can be almost certainly excluded 
because the stars would not be able to live as RSG long enough to undergo 
significant mass loss over there and perform short-lived loops at the end
of core He-burning. This is
particularly true for  low metallicities  for which 
evolution without overshoot leads more easily to the case B scheme. 
Case C models (whole core He-burning as RSG, and fully mixed overshoot
regions) are also very unlikely because in order to get extended loops
all the burden is left to mass loss during the RSG stages. 
Loops are unlikely to occur even with much higher rates of mass loss.

Therefore, case A models with the new mass-loss rates during the RSG
phase are best suited to performing extended loops,
and explaining the  many stars in  the BHG, and the enhancement in the surface
abundances of He and CNO elements seen in many of these objects. They are also
best suited to interpreting  the stellar content of individual
clusters, NGC~330 of the SMC as a prototype, in which  
several 
studies have revealed the presence of a significant number of blue
supergiant stars near the main sequence band 
(Caloi et al. 1993; Vallenari et al. 1994; Chiosi et al. 1995), 
which cannot be explained by other types of models.

\begin{figure*}
\centerline{\psfig{file=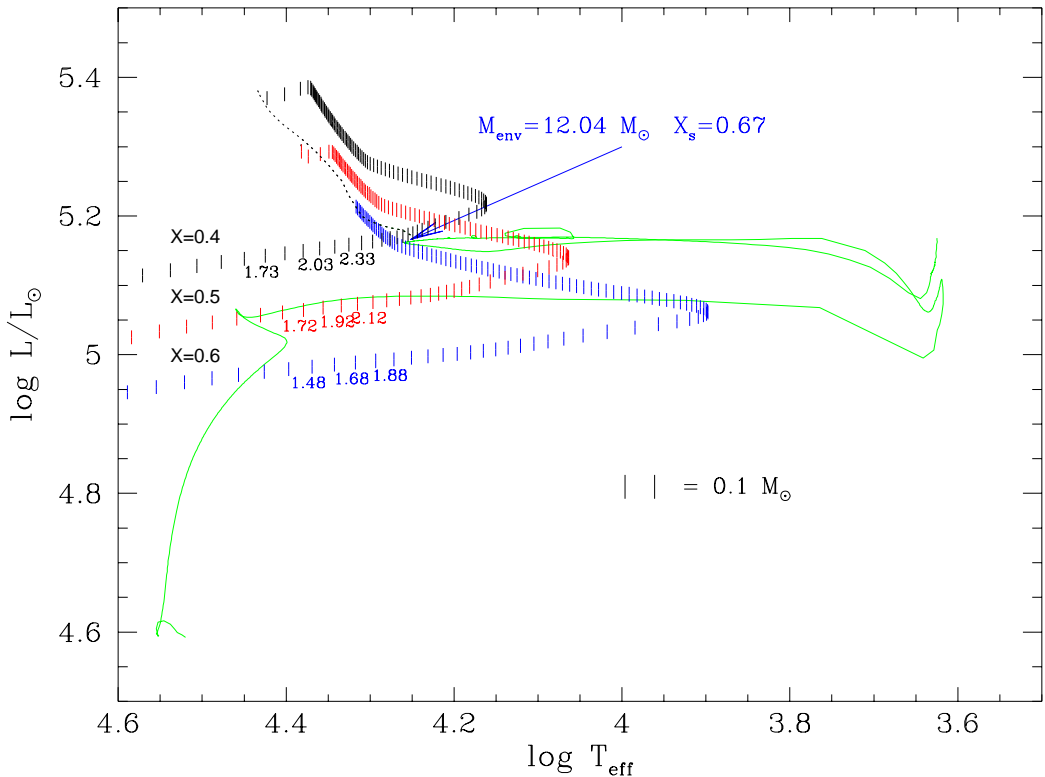,height=11.truecm,angle=270}}
\caption{Equilibrium models  at varying surface 
abundance of hydrogen and envelope mass. The solid line is the sequence of
20\msol\ model with [Z=0.008, Y=0.25] 
computed with our diffusive scheme and de Jager et al.
mass-loss rate. Starting from the bluest model of the loop, the
envelope H-abundance has been lowered to 0.6, 0.5 and 0.4
(the models evolve along the dotted line). At these selected values,
sequences of varying envelope mass have been constructed. The vertical
bars mark decrements of 0.1 \msol\ in the envelope mass. 
The numbers along the three sequences  indicate
the  envelope mass of the models.}
\label{enve}
\end{figure*}

However to fully understand the role of the internal mixing, in
particular whether it would be possible to recover  at least part 
of the results obtained with the much higher mass-loss rates in the red,
we have performed a series of numerical experiments on the evolutionary 
sequence of the case-A 20 \msol\  star calculated with the de
Jager et al. (1988)  mass-loss rates all over the HRD. 
 We anticipate that similar results can be
obtained using  the 20 \msol\ star calculated  with the fully mixed
overshoot scheme in the convective core and the Schwarzschild condition
for the instability in the intermediate convective shell. These latter
experiments are not shown here for the sake of brevity.

The 20 \msol\ star in usage here belongs to case-A evolution and when it leaves
the RSG phase the  envelope mass is about about 12 \msol. We take the
bluest model in  the loop as the reference model on which the experiments are
performed. Starting from this, we  calculate several sequences
of equilibrium models (without the gravitational energy source, 
cf. Lauterborn et al. 1971),  which have 
different surface  hydrogen  abundance  and envelope mass.
The adopted surface abundances of hydrogen are X=0.6, 0.5, and 0.4. 
The envelope mass is decreased in steps of 0.1 \msol. 
All these evolutionary sequences are shown in Fig.~\ref{enve}.

We first artificially decrease the envelope abundance to the
selected value. These  models  follow the dotted line in Fig~\ref{enve}.
Once the selected surface abundance has been reached,
we artificially decreased the envelope mass. Each vertical bar in Fig.
\ref{enve} marks  a decrement of 0.1 \msol. 

Decreasing the surface H-abundance makes the models hotter and
more luminous, because of the higher mean molecular weight. They move
toward a pure helium sequence.  
For instance, the starting models with X=0.5 and 0.4 even reach 
 the main sequence band of  more massive stars. 

In order to 
reach this configuration, an evolutionary scheme is required, in which 
an additional amount of about 2\msol\ (for X=0.5)  or  3\msol\ (for X=0.4) of 
hydrogen is burned and mixed
to the surface during or before the RSG phase. 
This could happen with a larger overshoot distance during the
main sequence phase (in order to increase the mass of the burned
hydrogen) followed by a deeper penetration of the convective envelope
during the RSG phase. However, this would imply that envelope
convection should be able to 
penetrate very deep inside a star  (up to several pressure scale heights). 
Unfortunately, the models computed by
Alongi et al. (1991) with envelopes whose overshooting distance is 0.7 
$H_P$, can actually  erode and mix only a tiny fraction of the inner 
He-rich core. Other mechanisms, such
as rotationally induced mixing, have been investigated by several
authors (cf. Eryurt et al. 1994, Talon et al. 1997). 
None of these models, however,   agrees with the kind of 
evolution suggested by the observational data.

The equilibrium sequences of decreasing envelope mass 
shown in Fig. \ref{enve} demonstrate  
that envelope masses in the range 1 to 2 \msol\ 
are required in order to populate the high effective temperature
region between the main sequence band and the loop. Clearly, the
higher the H-surface abundance, the lower is the envelope mass
at a given effective temperature.

Therefore, unless a still unknown mixing mechanism operates  to
lower the H-abundance in the envelope below 0.5, the only
possible way to populate the BHG is to get envelope masses below
 2 \msol. In this case internal mixing may help, but not cause, the
blueward extension of the model.

We conclude that only the combination of 
case-A evolution (which means 
a certain efficiency of mixing in the overshoot and intermediate
convective zones) with  the new  mass-loss rates during the RSG phase
can  solve the mystery of the many stars observed in the BHG.

It is worth noticing that  large mass-loss rates in the
 red have already been suggested, 
cf.  Chiosi et al. (1978)  who adopted a similar scheme. 
They made use of radiation-driven wind formulation during the main 
sequence phase, and adopted a mass-loss rate proportional to the acoustic 
flux in the
RSG phase. A mass-loss rate as high as few 10$^{-4}$\msol/yr was obtained
for their 20\msol\ model during the RSG phase. Unfortunately, due
to numerical difficulties their sequence  was not able to evolve
blueward of $\rm log T_{eff}\simeq 4. 2$, while the models were still
 performing the blue loop, so
that it was not possible to evaluate the maximum extension toward the blue. 
Because of  the high mass-loss rate applied during the RSG phase,
 their 20\msol\ model can be considered as the ancestor 
of the evolutionary scheme  suggested in this study.

\section{ Discussion and conclusions}

In this paper we have revisited the evolution of massive stars in the mass
range 15 to  30 \msol. This  mass range 
corresponds to the area of the HRD 
in which both blue and red supergiant stars coexist.   
Extant  observational data  provide hints
to understand some important mechanisms that are at the base of
the evolution of this class of stars. Despite  their relatively  simple
internal structure during both core H-  and He-burning, all the
models  computed  in the past fail in reproducing some well established
observational facts. First, the so called ``Blue
Hertzsprung Gap'', the region of the HRD near the main sequence
band which is well populated almost independently from the environment (Solar
Vicinity, LMC, SMC), and, on the contrary, is
crossed by the models in a Kelvin-Helmoltz time-scale. Second,
blue supergiant
stars in this area of the HRD are also He-enriched and show CNO
abundance characteristic of the first dredge-up episode, indicating
that they  underwent the RSG phase and are now burning  helium
in the center while performing a very extended blue loop.

To understand the reason of the failure of current models 
with respect to those observational constraints,  we re-examined the effects of
both internal mixing 
and mass loss by stellar winds (during the red stages in particular).

{\it Mixing.} Utilizing   suggestions from recent hydrodynamical
calculations,
we model internal mixing as a turbulent diffusive process and
account for the exponential decay of the convective velocities
outside the classical Schwarzschild border. The characteristic decay
scale is parameterized by a suitable fraction of the pressure scale
height ($\alpha_1$).

As far as the intermediate convective zone is concerned, we follow
Langer (1983, 1985) who included a time dependent
mixing process. The efficiency of this intermediate convective mixing
is parameterized by $\alpha_2$. 

By suitably choosing  the parameters $\alpha_1$ and $\alpha_2$
we can reproduce a wide range of mixing topologies, from the standard
semiconvective one to the full overshoot scheme.

We performed an extensive analysis  to determine the combination of
the parameters $\alpha_1$ and $\alpha_2$ yielding the most extended
loop for a typical 20\msol\ star  with LMC composition and 
mass loss by stellar wind according to the de Jager et al. (1988)
relationships.

The test experiments for the parameter selection and the 
corresponding  evolutionary models show that internal mixing alone cannot be
responsible of the observed star distribution in the region between
the main sequence band and the blue side of the loop. 

Despite the new prescription of internal mixing, these models are much 
similar to others in literature and share many of their weak points.
In brief, for the adopted values of $\alpha_1$ and $\alpha_2$, the
20 \msol\ star with LMC composition is able to perform extended loops (case A),
whereas
the same star but with solar composition has the whole core He-burning phase
as a RSG (case C). Finally, all these models predict the BHG in the HRD.\\

\begin{figure}
\psfig{file=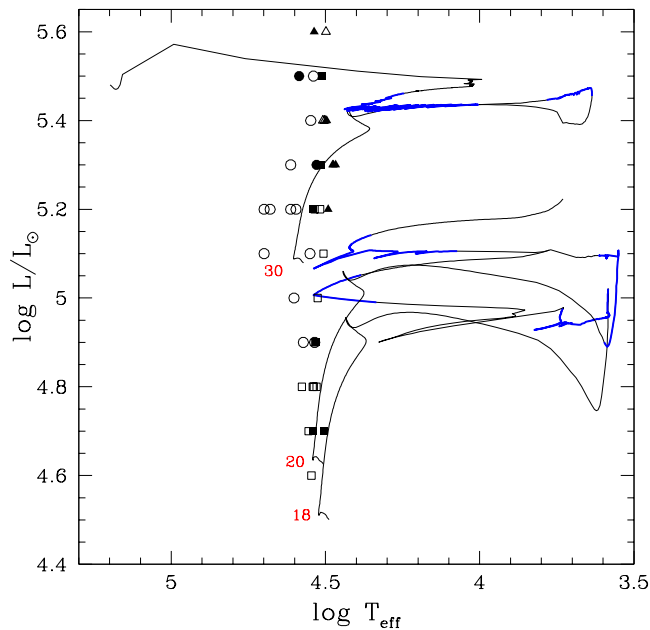,width=8.8cm}
\caption{HRD of the  18-20-30 \msol\ stars with 
composition [Z=0.02, Y=0.28], the new diffusive scheme, and the revised
 mass-loss rates during the RSG stages. The slow
stages of the core He-burning are indicated  by  heavier lines. 
Finally, the filled
and open circles show the WR stars of Hamann et al. (1993). The meaning of
the symbols is the same as in Fig.~8.  
Note the new channel for the formation of the 
low-luminosity WR stars (see the text for more details).}
\label{hrgrosso}
\end{figure}

{\it Mass loss}. Examining the  causes of the failure, we find that the
mass-loss rates in the RSG phase by de Jager et al. (1988) actually
 under-estimate the observational values  by a
large factor. 
Using the mass-loss rates  for red supergiants  by Feast
(1991), which are about an order of 
magnitude larger than the previous ones, the 
20\msol, 18\msol, and 15\msol\ stars with  solar composition (Z=0.020) 
perform  wide 
loops across the HRD up to the main sequence band. During the loop phase,
the models also possess the observed surface composition.
The reason for the extended loop is the small envelope mass 
($\leq$ 2\msol) in the star
after the RSG phase.
However, when the new rates are applied to the same star
with the  LMC composition
(Z=0.008) the loops are much narrower than expected.

This spurred us to revisit the mass-loss estimates upon which the 
Feast (1991) 
formulation is based (Jura \& Kleinmann 1990;  Reid et al.
1990). We find that the rates could be systematically under-estimated
as the constant dust to gas ratio
$\delta=0.0045$ typical of AGB stars was adopted. By suitable corrections 
for the physical conditions holding in red supergiant stars 
(different gas to dust ratio), we find that the
 mass-loss rates of Feast (1991) are 
likely under-estimated by a factor of five.
We remind the reader that even scaled by a factor of five the Feast (1991)
mass-loss rates are 
below the superwind regime, at which  the wind-momentum parallels  the
radiation-field momentum.

Because our goal is not to derive  new  estimates of the 
mass-loss estimate in the RSG
phase,  but  only to show that the mass-loss rates required in our
evolutionary scheme are  within  the observational uncertainty, 
 we  simply re-scale the  values given by Feast (1991), correct them for 
the likely variation of the dust to gas ratio,  and apply the new rates   to
stellar models.

Now  the 20\msol\ stars with  solar and LMC
composition  develop  extended loops
across the HRD.  
The same is true for the models of 15 and 18 \msol\ and LMC 
metallicity, and  finally for models of the same mass 
but the chemical composition typical of the  SMC (Z=0.004). 
They are  not described  here  for the  sake of brevity.\\

{\it Equilibrium models.} Finally, to clarify the combined 
role of mass-loss and internal
mixing we computed static equilibrium configurations at varying
surface hydrogen  abundance and envelope mass.
A   20\msol\ star with standard mass-loss rates in the red, 
while performing the loop, has  a H-rich envelope of
about 12 \msol.  If the H-abundance in the envelope of this star 
is let decrease by
some mixing mechanism down to 0.4, then the star would shift toward 
higher luminosity and hit the main sequence.
However, arguments are given to claim that an evolutionary 
scenario yielding this kind of stellar structure does not exist
and most likely is even not  possible.

In contrast,  static models  with conventional mixing but 
envelope mass below 2\msol\
populate the blue gap, naturally suggesting that mass-loss (in the red)
is the leading parameter of the problem.
The internal chemical profile
left out by the intermediate convective region  may somewhat affect
 but not cause the above trend.

Therefore,  we suggest that  a thorough, both observational and
theoretical revision of the mass-loss rates during  the RSG phase 
is urgently needed.\\

{\it WR stars.}
Finally, we like to draw some remarks on the use of these models to 
interpret the observational properties of WR stars.
We start noticing that soon after
leaving the RSG
region,  the surface H-abundance of the 18 and 20 \msol\ models
of solar composition is 0.43, i.e. very close to the value of 0.40
assumed by Maeder (1990) to mark the transition from  O-type  to  WNL
stars. Furthermore these values are  consistent, within the
uncertainty limits, with the ones observed by Hamann et al.
(1993) in some WNL stars.
Fig.~\ref{hrgrosso} shows the path in the HRD of the 18, 20
and 30 \msol\ stars with solar composition together with the
data for the WNL and WNE stars observed by Hamann et al. (1993).
There is an immediate new result coming out of this comparison,
i.e. a new avenue  for the
formation of faint single WR stars ($\rm log L/L_{\odot} \sim 4-4.5$).
Contrary to what suggested in the past, most likely the progenitors  of
the low luminosity WR stars   are not the massive stars 
(say $\rm M > 60$ \msol) evolving
``{\it vertically}'' in the HRD, but the less 
massive stars (say  18-20 \msol)  evolving ``{\it horizontally}''
provided they suffered from significant mass-loss during the RSG phase (cf.
Vanbeveren 1995). 
Soon after
the RGB stages, the models are structurally similar to  WNL stars,
but they become visible as such only near the main sequence, when the
effective temperature is high enough to show the spectral signatures
of  WR stars. Are the tiny  envelopes of these WR candidates 
unstable and prone to the high mass-loss
typical  of WR stars ? Answering this question deserves further
investigation.
\vspace{0.5cm}

{\it Acknowledgements.} The authors thanks financial support from the
Italian Ministry of University, Scientific Research and Technology
(MURST), the Italian Space Agency (ASI), and the European Community under 
TMR grant ERBFMRX-CT96-0086.


\begin{thebibliography}{}

\bibitem{} 
Abbott, D. C., 1982, ApJ, 259, 282

\bibitem{}
Alongi, M., Bertelli, G., Bressan, A.~G., \& Chiosi, C., 1991,
  A\&A,  244, 95

\bibitem{}
Alongi, M., Bertelli, G.~P., Bressan, A.~G., Fagotto, F., Greggio, L., 
      \&  Nasi,  E., 1993, A\&AS,  97, 851

\bibitem{}
Bertelli, G.~P., Bressan, A.~G., \& Chiosi, C., 1984, A\&A,  130, 279

\bibitem{}
Bowen, G.~H. \& Willson, L.~A., 1991, ApJL,  375, 53

\bibitem{}
Bressan, A., Granato, G.~L., \& Silva, L., 1997, A\&A, in press

\bibitem{}
Bressan, A.~G., 1994, Space Sci. Rev.,  66, 373

\bibitem{}
Bressan, A.~G., Bertelli, G.~P., \& Chiosi, C., 1981, A\&A,  102, 25

\bibitem{}
Bressan, A.~G., Fagotto, F., Bertelli, G., \& Chiosi, C., 1993,
  A\&AS,  100, 647

\bibitem{}
Caloi, V., Cassatella, A., Castellani, V., \& Walker, A., 1993,
  A\&A,  271, 109

\bibitem{}
Canuto, V.~M., 1992, ApJ,  392, 218

\bibitem{}
Canuto, V.~M., 1994, ApJ,  428, 729

\bibitem{}
Canuto, V.~M., 1996, ApJ,  467, 385

\bibitem{}
Canuto, V.~M. \& Mazzitelli, I., 1991,
  ApJ,  370, 295

\bibitem{}
Castor, J. I., Abbott, D. C., Klein, R. I., 1975, ApJ 195, 157

\bibitem{}
Charbonnel, C., Meynet, G., Maeder, A., Schaller, G., \& Schaerer, D., 1993,
A\&AS, 101, 415
 
\bibitem{} 
Chiosi, C., 1997,
in  {\it Stellar Astrophysics for the Local Group: a First Step to the
  Universe}, eds. A. J. Aparicio \& A. Herrero,  Cambridge University Press


\bibitem{}
Chiosi, C., Bertelli, G., \& Bressan, A.~G., 1992, ARA\&A,  30, 235

\bibitem{}
Chiosi, C. \& Maeder, G.~A., 1986, ARA\&A,  24, 329

\bibitem{}
Chiosi, C., Nasi, E., \& Sreenivasan, S.~R., 1978, A\&A,  63, 103

\bibitem{}
Chiosi, C. \& Summa, C., 1970, Astro. Space Sci.,  8, 478

\bibitem{}
Chiosi, C., Vallenari, A., Bressan, A., Deng, L., \& Ortolani, S., 1995,
  A\&A,  293, 710

\bibitem{}
de~Jager, C., Nieuwenhuijzen, H., \& van~der Hucht, K.~A., 1988,
  A\&AS,  72, 295

\bibitem{}
de~Koter, A., Heap, S., \& Hubeny, I., 1997, ApJ,  477, 792

\bibitem{}
Deng, L., 1993, Ph.D. Thesis, ISAS, Italy

\bibitem{}
Deng, L., Bressan, A., \& Chiosi, C., 1996, A\&A, 313, 145

\bibitem{}
Eryurt, D., Kirbiyik, H., Kiziloglu, N., Civelek, R., \& Weiss, A.,
 1994, A\&A, 282, 485

\bibitem{}
Fagotto, F., Bressan, A., Bertelli, G., \& Chiosi, C., 1994,
  A\&AS,  105, 39

\bibitem{}
Feast, M.~W., 1991,
  Instabilities in Evolved Super and Hyper-Giants,
  eds. C. de Jager \& H. Nieuwenhuijzen

\bibitem{}
Fitzpatrick, E.~L. \& Bohanna, 1993, ApJ,  404, 734

\bibitem{}
Freytag, B., Ludwig, H.~H., \& Steffen, M., 1996, A\&A,  313, 497

\bibitem{}
Gabriel M., 1995,
  in {\it Stellar Evolution: What Should Be Done}, 32th Liege International
  Astrophysical Colloquium, eds. Noels, A./ et al. 
  Liege: Inst. d'Astrophysique

\bibitem{}
Gail, H. P., Sedlmayr, E., 1987, A\&A, 171, 197

\bibitem{}
Grossman, S., 1996, MNRAS,  279, 305

\bibitem{}
Grossman, S. \& Narayan, R., 1993, ApJS,  89, 361

\bibitem{}
Grossman, S., Narayan, R., \& Arnett, D., 1993, ApJ,  407, 284

\bibitem{}
Grossman, S.~A. \& Taam, R.~E., 1996, MNRAS,  283, 1165

\bibitem{}
Habing, H.~J., Tignon, J., \& Tielens, A. G. G.~M., 1994, A\&A,  286, 523

\bibitem{}
Hamann, W.~R., Koesterke, L., \& Wessolowski, U., 1993, A\&A,  274, 397

\bibitem{}
Heger, A., Jeannin, L., Langer, N., \& Baraffe, I., 1997,
  astro-ph/,  9705097, to appear in A\&A

\bibitem{}
Herrero, A., Kudritzki, R.P., , Vilchez, J, Kunze, D., 
Butler, K. \& Haser, S., 1992,
  A\&A,  261, 209

\bibitem{}
Herwig, F., Blocker, T., Schonberner, D., \& Eid, M.~E., 1997,
  A\&A,  324, 81

\bibitem{}
Hillier, D.~J., 1996,
  in U.~H. S.~Jeffery (ed.), Hydrogen Deficient Stars, 2nd
  International Colloquium, ASP Conf.\ Ser.

\bibitem{}
Ivezic, Z. \& Elitzur, M., 1995, ApJ,  445, 415

\bibitem{}
Jura, M., 1986, ApJ,  303, 327

\bibitem{}
Jura, M. \& Kleinmann, S.~G., 1990, ApJS,  73, 769

\bibitem{}
Kato, S., 1966, PASJ,  18, 374

\bibitem{}
Kudritzki, R.~P., 1997, 
in  {\it Stellar Astrophysics for the Local Group: a First Step to the
  Universe}, eds. A. J. Aparicio \& A. Herrero,  Cambridge University Press

\bibitem{}
Kudritzki, R.~P., Gabler, R., Groth, H.~G., Pauldrach, A.~W., \& Puls, J.,
  1989,
  in IAU Colloquium 113 p. 67

\bibitem{}
Kwok, S., 1975, ApJ, 1908, 583

\bibitem{}
Lafon, J. P.J., Berruyer, N., 1991, A\&A Rev. 2, 249

\bibitem{}
Lamers, H. \& Cassinelli, J., 1996,
  ASP Conf. Ser.,  98, 162

\bibitem{}
Langer, N., 1989,  A\&A,  210, 93

\bibitem{}
Langer, N., 1991,  A\&A,  252, 669

\bibitem{}
Langer, N., Eid, M. F.~E., \& Baraffe, I., 1989,  A\&A,  224, 17

\bibitem{}
Langer, N., Eid, M. F.~E., \& Frike, K.~J., 1985,  A\&A,  145, 179

\bibitem{}
Langer, N., Sugimoto, D., \& Frike, K.~J., 1983,  A\&A,  126, 207

\bibitem{}
Lauterborn, D., Refsdal, S., \& Weigert, A., 1971,  A\&A,  10, 97

\bibitem{}
Lennon, D.J., Dufton, P.L., Mazzali, P. A., Pasian, F., \& Marconi, G.
A\&A, 314, 243

\bibitem{}
Maeder, A., 1990,  A\&AS,  84, 139

\bibitem{} 
Maeder, A., \& Conti, P. S., 1994, ARA\&A, 32, 227

\bibitem{}
Maeder, A. \& Meynet, G., 1994,  A\&A,  287, 803

\bibitem{}
Massey, P., 1997,
in  Stellar Astrophysics for the Local Group: a First Step to the
  Universe, eds. A. J. Aparicio \& A. Herrero,  Cambridge University Press

\bibitem{}
Massey, P., Armandroff, T., \& Pyke, R., 1995a,  AJ,  110, 2715

\bibitem{}
Massey, P., Lang, C.~C., DeGioia-Eastwood, K., \& Garmany, C., 1995b,
  ApJ,  438, 188

\bibitem{}
Merryfield, W., 1995,  ApJ,  444, 318

\bibitem{}
Meynet, G., Maeder, A., Schaller, G., Schaerer, D., \& Charbonnel, C., 1994,
  A\&AS,  103, 97

\bibitem{}
Pijpers, F. P., \& Habing H. J., 1989, A\&A, 215, 334

\bibitem{}
Pijpers, F. P., \& Hearn, A. G., 1989, A\&A, 209, 198

\bibitem{}
Owocki S. P., Castro, J. I., \& Rybicki, G. B. 1988, ApJ, 335, 914

\bibitem{}
Pauldrach, A. Puls, J., \& Kudritzki, R. P., 1986, A\&A, 164, 86

\bibitem{}
Reid, N., Tinney, C., \& Mould, J., 1990,  ApJ,  348, 98

\bibitem{}
Schaerer, D., 1996,  A\&A,  309, 129

\bibitem{}
Schaerer, D., Charbonnel, C., Meynet, G., Maeder, A., Schaller, G., 1993,
A\&AS 102, 339

\bibitem{}
Schwarzschild, M. \& Harm, R., 1965,  ApJ,  142, 855

\bibitem{}
Talon, S., Zahn, J.P., Maeder, A., \& Meynet, G., 1997, A\&A, 322, 209

\bibitem{}
Vallenari, A., Ortolani, S., \& Chiosi, C., 1994,  A\&AS,  571, 108

\bibitem{}
Vanbeveren, D., 1995, A\&A, 294, 107

\bibitem{}
Vassiliadis, E. \& Wood, P.~R., 1993,  ApJ,  413, 641

\bibitem{}
Venn, K.A., 1995, ApJ, 449, 838

\bibitem{}
Waever, T.~A. \& Woosley, G.~B., 1978,  ApJ,  225, 1021

\bibitem{}
Willson, L. A. 1988, in Pulsation and mass loss in Stars, eds. R. Stalio \&
L.A. Willson, (Dordrecht: Reidel), p. 285


\bibitem{}
Willson, L.~A., Bowen, G.~H., \& Struck, C., 1995,
  Bull. Am. Astron. Soc.,  187, 103

\bibitem{}
Xiong, D.~R., 1979,  Acta Astron. Sin.,  20, 238

\bibitem{}
Xiong, D.~R., 1984, Sci. Sinica,  24, 1406

\bibitem{}
Xiong, D.~R., 1985,  A\&A,  150, 133

\bibitem{}
Xiong, D.~R., 1989,  A\&A,  213, 176

\bibitem{}
Zahn, J.~P., 1991,  A\&A,  252, 179

\end{thebibliography}
\end{document}